\pdfoutput=1
\documentclass[lettersize,journal]{IEEEtran}
\usepackage{amsmath,amsfonts}
\usepackage[linesnumbered,ruled]{algorithm2e}
\usepackage{array}
\usepackage[caption=false,font=normalsize,labelfont=sf,textfont=sf]{subfig}
\usepackage{textcomp}
\usepackage{threeparttable}
\usepackage{longtable}
\usepackage{booktabs}
\usepackage{stfloats}
\usepackage{url}
\usepackage{verbatim}
\usepackage{graphicx}
\usepackage{cite}
\usepackage{multirow}

\begin{document}

% content

\title{The Hybrid ROA: A Flexible and Scalable Encoding Scheme for Route Origin Authorization\\
%{\footnotesize \textsuperscript{*}Note: Sub-titles are not captured in Xplore and
%should not be used}
%\thanks{Identify applicable funding agency here. If none, delete this.}
}

%add author and affi. later
\author{
\IEEEauthorblockN{
Yanbiao Li\IEEEauthorrefmark{1}\IEEEauthorrefmark{2}, 
Hui Zou\IEEEauthorrefmark{1}, 
Yuxuan Chen\IEEEauthorrefmark{1}\IEEEauthorrefmark{2}, 
Yinbo Xu\IEEEauthorrefmark{1}\IEEEauthorrefmark{2}, 
Zhuoran Ma\IEEEauthorrefmark{1}\IEEEauthorrefmark{2}, 
Di Ma\IEEEauthorrefmark{3}, 
Ying Hu\IEEEauthorrefmark{1} and 
Gaogang Xie\IEEEauthorrefmark{1}\IEEEauthorrefmark{2}}
\IEEEauthorblockA{\IEEEauthorrefmark{1} 
\textit{Computer Network Information Center},
\textit{Chinese Academy of Sciences}, Beijing, China}
\IEEEauthorblockA{\IEEEauthorrefmark{2} 
\textit{School of Computer Science and Technology},
\textit{University of Chinese Academy of Sciences}, Beijing, China}
\IEEEauthorblockA{\IEEEauthorrefmark{3} 
\textit{Internet Domain Name System Beijing Engineering Research Center (ZDNS)}, Beijing, China}
}

\maketitle
\begin{abstract}%0.25p
On top of the Resource Public Key Infrastructure (RPKI), the Route Origin Authorization (ROA) creates a cryptographically verifiable binding of an autonomous system to a set of IP prefixes it is authorized to originate. By their design, ROAs can protect the inter-domain routing system against prefix and sub-prefix hijacks. However, it is hard for the state-of-the-art approach, the maxLength-based ROA encoding scheme, to guarantee security and scalability at the same time when facing various authorization scenarios. To this end, we propose a novel bitmap-based encoding scheme for ROAs to provide flexible and controllable compression. Furthermore, the hybrid ROA encoding scheme (\texttt{h-ROA}) is proposed, which encodes ROAs based on \texttt{maxLength} and \texttt{bitmap} jointly. This approach ensures strong security, provides flexibility and significantly improves system scalability, enabling it to effectively handle various authorization patterns. According to the performance evaluation with real-world data sets, \texttt{h-ROA} outperforms the state-of-the-art approach $1.99 \sim 3.28$ times in terms of the encoding speed, and it can reduce the cost of a router to synchronize all validated ROA payloads by $43.9\% \sim 56.6\%$.

%inappropriate configurations bring in vulnerabilities to other types of routing security attacks. As such, the state-of-the-art approach implements the minimal-ROA principle, eliminating the risk of using ROAs at the cost of system scalability. This paper proposes the hanging ROA, a novel bitmap-based encoding scheme for ROAs, that not only ensures strong security, but also significantly improves system scalability. According to the performance evaluation with real-world data sets, the hanging ROA outperforms the state-of-the-art approach $2.4$ times in terms of the compression ratio, and it can reduce the cost of a router to synchronize all validated ROA payloads by $44.5\% \sim 64.7\%$.
\end{abstract}

\section{Introduction}%1p
\label{sect:intro}

\subsection{Border Gateway Protocol and Prefix Hijacking}
The Border Gateway Protocol (BGP)~\cite{rfc1105-bgp, rfc4271-bgp4} is one of the fundamental building blocks of the global Internet, and is designed for Autonomous Systems (ASes) to exchange routing and reachability information. However, its basic assumption that every component tells the truth about its reachability does not always hold in reality. This is the root cause to some of the most devastating attacks towards the inter-domain routing system with BGP, such as \emph{prefix hijacks}~\cite{prefix-hijacks-1, artemis}. For example, by deceiving others into believing that it is the origin of some IP prefix(es) belonging to an AS, the attacker can steal all or part of the network traffic heading to this AS, causing many victims to have their access disrupted and, in more serious cases, their data stolen. These attacks are devastating and propagate quickly, and are easy to launch but hard to detect. 

\subsection{Resource Public Key Infrastructure}
Securing the inter-domain routing system with BGP requires great attention, and has been studied for decades. The Resource Public Key Infrastructure (RPKI)~\cite{rfc6480-rpki, rfc8210-rpki} is the only approach that has seen widespread deployment so far~\cite{nist-monitor}. It aims at enhancing the security of BGP with proactive protections, by cryptographically certifying the ownership and usufruct of Internet number resources, such as IP addresses.

In RPKI, there are three major kinds of entities. The owner of a piece of Internet number resource, regarded as a \emph{Certificate Authority} (CA), holds the resource certificate and can further distribute the right to use the resource to others via cryptographically verifiable data objects. Then, with these objects, an RPKI-enabled \emph{Router} can validate BGP messages and make routing decisions per their validity status. This can help the router to stop potential BGP security attacks at the earliest stage~\cite{router-use-rpki, miller2019taxonomy}. However, it can be a great burden for a router to synchronize and verify RPKI objects. Accordingly, a Relying Party (RP) acts as the agent of one or more routers to synchronize RPKI objects, from a distributed repository where CAs publish their data, and using the RPKI Repository Delta Protocol (RRDP)\footnote{Some CAs and RPs may use its predecessor \texttt{rsync}.}~\cite{rfc8182-rrdp}. In addition, the RP stores verified RPKI objects with a local cache, where the router(s) can fetch them using the RPKI to Router (RTR) protocol~\cite{rfc6810-rtr}. 

% why ROA
On top of the RPKI, several approaches have been proposed to secure the inter-domain routing system with BGP~\cite{rfc8206-bgpsec, rfc8374-bgpsec, li-bgpsec, decentral-bgpsec, aspa-profile, aspa-verification}. However, most of them are confronted with great resistances in practical deployment, as they either bring in heavyweight online cryptography (like BGPSec~\cite{rfc8206-bgpsec, li-bgpsec}), or demand sensitive business information (such as the ASPA~\cite{aspa-profile}). To date, the Route Origin Validation (ROV) with Route Origin Authorizations (ROAs)~\cite{rfc6482-roa, rfc6483-rov, rfc6811-rov, rfc7115-rov} is the only approach that has finished its standardization and is widely deployed~\cite{nist-monitor}. The ROA is an RPKI object that certifies the authorization of an AS to originate a set of IP prefixes. As such, the use of ROAs can effectively stop prefix hijacks~\cite{roa-hijack,clark2020filter}.

\subsection{Challenges with ROA Encoding}
Though ROAs are useful in securing BGP, it is challenging to use them efficiently. The encoding of ROAs is confronted with four key challenges.

\subsubsection{Compression}
Regarding a global deployment, it can take a long time to transmit data from the producers (CAs) of ROAs to their consumers (routers)~\cite{rpki-full-size}. The faster the router fetches the latest ROAs, the earlier it gets the timely knowledge for security protection. Thus, ROAs should be encoded with compression to improve the efficiency of transmission and processing. This is important to reduce burdens on routers when ROV with ROAs is enabled.

\subsubsection{Security}
As the use of ROAs is for security, the encoding scheme itself must be secure enough. Otherwise, the validation using ROAs becomes unreliable and thus useless. More specifically, the encoding semantic should be meaningful and accurate, without bringing in any potential security hazards, especially when deep compression is made.

\subsubsection{Flexibility}The quantity and distribution pattern of authorized IP prefixes varies among different ASes, which is ever-changing. Therefore, it is important for the ROA encoding scheme to be sufficiently flexible to adapt to different authorization patterns, such as scattered or concentrated, in order to achieve maximum compression efficiency.

\subsubsection{Scalability}
The number of IP prefixes to authorize with all ASes determines the number and size of ROAs, which directly or indirectly affect the cost of synchronization, verification and maintenance. Today, around $50\%$ of the routes in the global Internet have been covered with ROAs, and this percentage is increasing fast~\cite{nist-monitor}. The future is coming! In addition, the number of routes increases year by year as well~\cite{bgp-reports}. Accordingly, regarding a full deployment in near future~\cite{rpki-full-size, rpki-study, rpki-coming}, the encoding scheme of ROAs must scale well with the number of IP prefixes to authorize.

\subsection{Limitation of Prior Arts}
\label{sect:intro:prior}

% rfc6482, maxlength; 7115 conservative use of maxlength
The traditional solution~\cite{rfc6482-roa}, which has finished its standardization and been deployed in today's RPKI system, encodes an ROA with an AS number (ASN) and a sequence of \texttt{ipAddrBlocks}, each containing an IP address prefix and an optional field \texttt{maxLength}. It specifies the maximum length of the IP address prefix that the AS is authorized to originate. For example, if an ROA authorizes a certain AS to originate \texttt{202.127.16.0/20} and the \texttt{maxLength} is set to 21, this AS is actually authorized to originate three IP address prefixes: \texttt{202.127.16.0/20}, \texttt{202.127.16.0/21}, and \texttt{202.127.24.0/21}. The use of \texttt{maxLength} enables ROA compression by packing several IP prefixes into one block, and offers the operators the convienience for route de-aggregation. We call this approach the \texttt{t-ROA} in short.

Nevertheless, extensive measurements of the deployed RPKI system have revealed that the use of \texttt{maxLength} is vulnerable to forged-origin sub-prefix hijacks~\cite{forged-origin-attack}. As such, it is recommended to use this field conservatively~\cite{rfc7115-rov}. A straightforward solution is to eliminate the \texttt{maxLength} parameter by authorizing a set of IP prefixes one by one~\cite{m-roa}. We call this approach the \texttt{s-ROA} in short. Obviously, the number of \texttt{ipAddrBlocks} in an ROA will increase sharply, and so will the number of RTR payload Protocol Data Units (PDUs) with each containing a tuple (IP prefix, maximum length, ASN), whose prefix length equals to maximum length. The more the PDUs, the higher the burden to the router for synchronization and processing. Further, the latest approach~\cite{m-roa, m-roa-rfc} argues that many of the benefits came with \texttt{maxLength} can be achieved without exposing users to attacks, and recommends to keep the \texttt{minimal ROA} principle whenever possible, that only those prefixes actually authorized to advertise with BGP are encoded with ROAs. To this end, a compression algorithm has been proposed~\cite{m-roa} to compress a list of PDUs that do not use excessive \texttt{maxLength} to the one that does, provided that all IP address prefixes covered by the produced PDU are exactly the ones carried with previous PDUs. Though this can achieve compression to some extent, the condition to compress is restrictive. We call this approach the \texttt{m-ROA} in short.

To sum up, \texttt{t-ROA}, \texttt{s-ROA} and \texttt{m-ROA} are essentially maxLength-based ROA encoding schemes (\texttt{ML-ROA} in short), and the difference between them lies in how they exploit \texttt{maxLength} for compression. More specifically, regarding the four challenges with encoding ROAs mentioned above, the \texttt{t-ROA} achieves a deep compression through the excessive use of \texttt{maxLength}, but it brings security risks because it authorizes IP prefixes that should not be authorized, unless authorized IP prefixes of an AS are concentrated enough to form a complete binary tree; the \texttt{s-ROA} offers flexibility but abandons the compression capability provided by \texttt{maxLength}, resulting in unacceptable scalability; the \texttt{m-ROA} ensures security by using \texttt{maxLength} with a ``best-effort" compression, but actually restricts the degree of compression and thus scalability of the system. Besides, \texttt{m-ROA} can provide better flexibility than \texttt{t-ROA} when facing the scattered authorized IP prefixes, as illustrated in Figure~\ref{fig:example}, albeit at the expense of scalability. 
%In this paper, \texttt{m-ROA} is selected as the representative of the maxLength-based ROA encoding scheme, which provides a certain degree of compression, scalability and flexibility without compromising security.

\subsection{Our Contributions}
\label{sect:intro:contribution}

In this paper, we present the design of the bitmap-based ROA encoding scheme (\texttt{BM-ROA} in short), which utilizes bitmaps to precisely manage authorized IP prefixes, thereby providing flexible and controllable compression. Furthermore, we propose the hybrid ROA scheme (\texttt{h-ROA} in short) that combines \texttt{BM-ROA} and \texttt{ML-ROA} encoding for a mix of cases involving scattered and concentrated authorized IP prefixes, to achieve the maximum compression efficiency. \texttt{h-ROA} is as secure as \texttt{m-ROA} and \texttt{s-ROA}, but is more flexible and scalable than \texttt{m-ROA}, \texttt{s-ROA} and \texttt{t-ROA}. Overall, we make the following contributions.

\begin{itemize}
    \item In Section~\ref{sect:hroa}, we propose \textit{bitmap-based ROA} (\texttt{BM-ROA}) to concisely and accurately manage a set of IP address prefixes authorized to originate from an AS as a sub-tree of the IP address trie, with a \texttt{bitmap} to effectively track and verify the states of individual IP address prefixes.
    
    \item We then propose the use of hanging levels to split the IP prefix trie into non-overlapping sub-trees, so as to balance the number and the size of bitmaps produced for a given ROA. In Section~\ref{sect:opt}, we also study in depth how to optimize the setting of hanging levels to maximize the compression.

    \item In Section~\ref{sec:hybrid}, we further propose the hybrid ROA (\texttt{h-ROA}), which flexibly chooses between \texttt{bitmap} and \texttt{maxLength} for encoding based on the distribution pattern of the authorized IP prefixes of an AS, to achieve maximum compression effect when scattered and concentrated authorized cases coexist.
    
    \item In Section~\ref{sect:system}, we prototype our scheme with the minimal modifications of existing protocols and software packages, and deploy it with open-sourced software packages for both the RP and the router.
    
    \item In Section~\ref{sect:exp}, we conduct extensive experiments to compare the performance of \texttt{h-ROA} with that of \texttt{t-ROA} and \texttt{m-ROA} using traces captured from the operational RPKI system and data sets collected in view of RPKI's full deployment. The results demonstrate clearly that \texttt{h-ROA} can achieve deeper compression and higher scalability. In an end-to-end system evaluation with the RTR \texttt{reset} query between the router and the RP, \texttt{h-ROA} reduces the overall cost on the router by at most $47.1\%$ and $64.7\%$, in comparison with \texttt{t-ROA} and \texttt{m-ROA} respectively.
    
\end{itemize}

\section{The Bitmap-based ROA Scheme}%2.5p
\label{sect:hroa}

\subsection{Terminology}
\label{sect:hroa:term}
To begin with, we define and clarify some important terms used throughout this paper.

We refer to a subtree of the IP address trie with the height 33 and 129 for IPv4 and IPv6 respectively as a \emph{sub-tree} shortly. Every node of the IP address trie\cite{trie} uniquely represents an IP address prefix, which is called the \emph{node's prefix} in short. We define an \emph{authorized prefix} of an AS as the IP address prefix authorized for the AS to originate. The least and most significant bits of an integer are referred to as LSB and MSB respectively.

As defined in~\cite{rfc6482-roa}, an ROA contains an ASN and a sequence of \texttt{ipAddrBlocks}, each encoding a tuple (IP prefix, maximum length). We call such a tuple an \emph{address block}. In RTR~\cite{rfc6810-rtr}, a validated ROA produces a list of PDUs, each associating the origin ASN of the ROA with one of its address blocks. As such, the number and the size of PDUs produced by a validated ROA reflect the compression effect of an ROA encoding scheme, and determines the cost on a router to fetch and use this ROA.

\subsection{The motivation example and key idea}% 1 pic, 0.25p
\label{sect:hroa:example}

\begin{figure}[tbp]
    \centering
    \includegraphics[width=\linewidth]{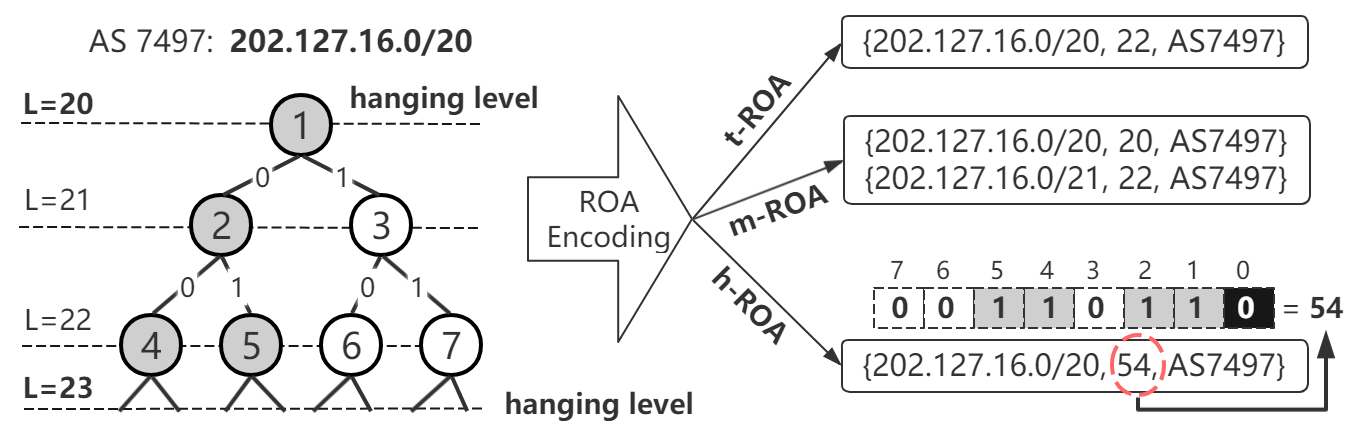}
    \caption{ROA encoding: AS 7497 is authorized to originate 4 IP prefixes.}
    \label{fig:example}
\end{figure}

We use the running case in Fig.~\ref{fig:example} as an example to illustrate our motivation and the key idea. In this example, \texttt{AS 7497} is authorized to originate 4 IP prefixes: \texttt{202.127.16.0/20}, \texttt{202.127.16.0/21}, \texttt{202.127.16.0/22} and \texttt{202.127.20.0/22}. Of the IPv4 prefix trie, a sub-tree rooted at the node representing \texttt{202.127.16.0/20} covers these 4 IP prefixes.

With \texttt{t-ROA}, only one PDU is produced to encode the whole sub-tree by setting the \texttt{maxLength} of the address block to 22. Unfortunately, this approach opens an attack vector to the forged-origin sub-prefix hijack~\cite{m-roa}, since some sub-prefixes, such as \texttt{202.127.24.0/21} represented by the node 3 of the sub-tree, will be treated as authorized prefixes but they are not. To avoid this risk, \texttt{m-ROA} keeps the principle of attesting only the prefixes that are indeed authorized. Namely, a PDU encodes a whole sub-tree if and only if all prefixes covered by this sub-tree are authorized. Accordingly, in this example, two PDUs are produced with \texttt{m-ROA}. Although \texttt{m-ROA} offers a secure solution to ROA encoding, its scalability is lower than \texttt{t-ROA}, because it may have to produce an identical PDU for every authorized prefix, in case there is no sub-tree satisfying its compression condition, for example, \texttt{AS 7497} is authorized to originate 3 IP prefixes: \texttt{202.127.16.0/20}, \texttt{202.127.16.0/21} and  \texttt{202.127.16.0/22}. 

From this, it can be seen that extensively scattered authorized IP prefixes of an AS will cause the \texttt{maxLength} in \texttt{m-ROA} to be ineffective for compression, degrading into \texttt{s-ROA} and compromising scalability.
To address this, we propose the design of \texttt{BM-ROA} for both secure and scalable ROA encoding, especially when authorized IP prefixes are scattered. The key idea is novel yet straightforward. With \texttt{BM-ROA}, the \texttt{maxLength} field is replaced by a \texttt{bitmap} encoding the whole sub-tree. With such a \texttt{bitmap}, every prefix covered by the sub-tree is represented by one identical bit, whose value claims whether this prefix is authorized (`1') or not (`0'). As such, \texttt{ BM-ROA} is as secure as \texttt{m-ROA}, because any unauthorized prefix can be easily identified. In addition, it offers even more flexibility than \texttt{t-ROA} in compression, by eliminating the de-facto limit with \texttt{t-ROA} that the root of an encoded sub-tree must represent an authorized prefix. In this example, only one PDU is produced to encode the whole sub-tree into a \texttt{bitmap}, where the bits corresponding to the 4 authorized prefixes are set.

\subsection{The scatter degree of authorized IP prefixes for each AS}
\label{sect:hroa:scatter}
Whether the authorized IP prefixes of an AS are scattered or concentrated largely depends on the principles a CA follows for authorizing IP prefixes, and this affects the compression effectiveness of \texttt{m-ROA}. A CA can follow the \texttt{minimal ROA} principle for security considerations, authorizing only the IP prefixes that need to be advertised in BGP; it can also, for the sake of convenience, authorize an IP prefix and all its subprefixes with the prefix length not exceeding specified \texttt{maxLength}. Intuitively, authorized IP prefixes issued following the \texttt{minimal ROA} principle are more scattered, whereas the opposite is true for those that do not. For this purpose, we define the \textit{scatter degree} of the authorized IP prefixes of an AS as the ratio of the number of complete binary trees they can form to the number of authorized IP prefixes, and use this metric to qualify compression capability of \texttt{m-ROA}. The larger the \textit{scatter degree} value of the authorized IP prefixes of an AS, the more entries are encoded, the weaker the ability of \texttt{m-ROA} to compress these authorized IP prefixes.

\begin{figure}[!tbp] 
  \begin{minipage}[b]{0.48\linewidth}
    \centering
    \includegraphics[width=\textwidth]{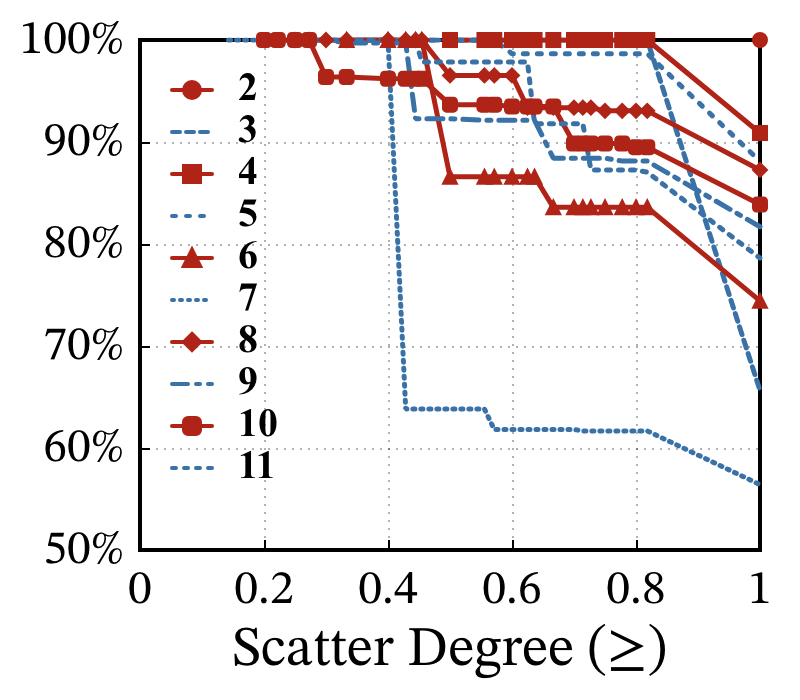}
    \caption{Scatter degree of \texttt{curr}.}\label{hybrid:degree:curr}
  \end{minipage}%
  \hfill
  \begin{minipage}[b]{0.48\linewidth}
    \centering
    \includegraphics[width=\textwidth]{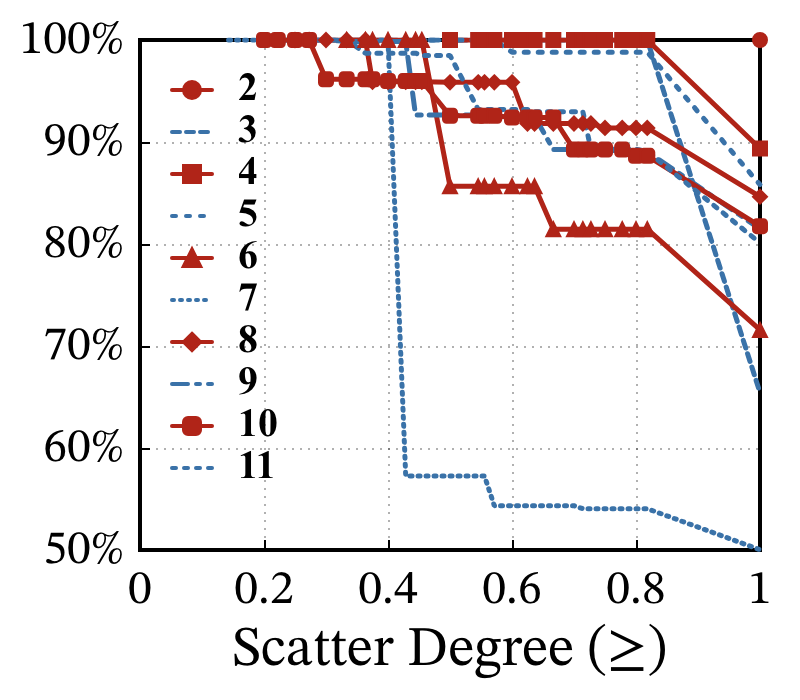}
    \caption{Scatter degree of \texttt{full}.}\label{hybrid:degree:full}
  \end{minipage}
\end{figure}

We calculate the \textit{scatter degree} of the authorized IP prefixes of each AS (except for AS0) that adhere to the \texttt{minimal ROA} principle by using the minimal ROAs of the data set \texttt{curr} and \texttt{full} (in Section~\ref{sect:exp:data}) that correspond to the current practice of
RPKI and its full deployment in the future respectively. For each set of minimal ROAs, we categorize ASes based on the number of IP prefixes that are authorized to them. Then, we select the top 10 groups with the most authorized IP prefixes, except for the groups with only one authorized IP prefix, which have no necessity for encoding and plot the CCDF of the scatter degree of the authorized IP prefixes of each AS within each group in Fig.~\ref{hybrid:degree:curr} and ~\ref{hybrid:degree:full}. The scatter degree of 1 indicates that the number of encoded entries is equal to the number of authorized IP prefixes, which implies that \texttt{maxLength} can not achieve any compression, and the proportion of ASes whose authorized IP prefixes with a scatter degree of 1 in each group is at least half. Additionally, in both \texttt{curr} and \texttt{full}, the number of ASes authorized to originate two IP prefixes peaks; however, the scatter degree of all ASes in this group equates to 1, meaning that they are completely non-compressible via \texttt{maxLength}.

% 说明scattered authorization是广泛存在的；

\subsection{Basic encoding scheme}% 2 pics, 0.75p
\label{sect:hroa:basic}
We propose to encode a set of authorized prefixes of an AS into two parts. A sub-tree covering these prefixes is encoded into a \emph{bitmap}, while the prefix represented by the root of this sub-tree is encoded as its \emph{identifier}. In addition, we associate with a sub-tree a \emph{withdrawal flag}, indicating whether the authorization of all authorized prefixes covered by this sub-tree should be withdrawn.
 
\subsubsection{Encoded sub-tree}
Given a sub-tree covering a set of authorized prefixes of an AS, we propose to encode it into a bitmap of $2^{h}$ bits, where $h$ is the height of this sub-tree. In this bitmap, the LSB encodes with the withdrawal flag, and by default is 0. Each of the rest bits represents an identical node of this sub-tree. By numbering all tree nodes from 1 and following the level order traversal, every node can locate one bit of the bitmap using its node number as the index, starting from the bit next to the LSB. Except the LSB, any bit of the bitmap is set if and only if its corresponding node of the sub-tree represents an authorized prefix.

Here comes the key design issue: \emph{how does one encode a sub-tree that has a large height?} 
It is unwise to encode such a ``tall" tree directly, because the length of the bitmap increases exponentially with the height of the tree to encode. For instance, it requires $2^{33}$ bits to encode the whole IPv4 trie. Our solution is to divide a ``tall" tree into small parts and encode them separately. By specifying a sequence of hanging levels on the IP address trie, it is divided into a set of \textbf{non-overlapping} sub-trees, each rooted at a hanging level. A sub-tree rooted at the hanging level $i$ and terminates at the next hanging level $j$ (not included) has a height $j-i$,  and can be encoded with a bitmap of $2^{j-i}$ bits.
%\rev{We say that a node at level-$i$ can hang to level-$j$ when level-$j$ is a hanging level and $i \ge j$. Then, any node of the IP address trie belongs to an unique sub-tree, which is rooted at the nearest hanging level it can hang to.} Given two successive hanging levels level-$i$ and level-$j$, the height of a sub-tree rooted at level-$i$ must be \rev{$j-i$},  which can be encoded into a bitmap of $2^{j-i}$ bits. 
In Fig.~\ref{fig:example}, 20 and 23 are the root and leaf hanging levels of the sub-tree shadowed with dark gray, the sub-tree can be encoded into a bitmap of $2^{23-20}=8$ bits. Since the node-$1$, node-$2$, node-$4$, and node-$5$ of this sub-tree represent authorized prefixes, the bitmap encoding this sub-tree should have the bit-$1$, bit-$2$, bit-$4$ and bit-$5$ set accordingly.

 \subsubsection{Sub-tree identifier}
 With the specification of hanging levels, any sub-tree is rooted at a hanging level, and ends by the one before the next hanging level or the last tree level. Namely, a node on a hanging level uniquely defines a sub-tree rooted at it, whose prefix can be encoded as the identifier of this sub-tree as follows. Suppose the depth of this hanging level is $l$ (starting from 0) - in this case, the prefix of this sub-tree's root contains exactly $l$ bits. We propose to encode this prefix into a $(l+1)$-bit integer as the sub-tree's identifier, by concatenating a set bit and all bits contained in this prefix. This scheme ensures the uniqueness of the identifier, because the position of the most significant set bit of an encoded prefix determines its prefix length. Take the sub-tree in Fig.~\ref{fig:example} as an example. \texttt{202.127.16.0/20} is the prefix represented by its root and the hanging level is 20. Then, the identifier of this sub-tree is calculated as 1878001 (\textbf{1}11001010011111110001 in binary), which is made by concatenating a bit `1' and the first 20 bits of this prefix.
 
\subsubsection{The bitmap-based ROA}
In our approach, all authorized prefixes of an AS are divided into several sub-sets, each is maintained with an encoded sub-tree. As such, a bitmap-based ROA encodes an ASN and a list of \emph{sub-tree blocks}, each containing an encoded sub-tree and its identifier. As for an RRDP response, a bitmap-based ROA is encapsulated into its payloads directly. In RTR, a bitmap-based ROA will produce a list of PDUs, each containing a sub-tree block along with the origin ASN.

\subsection{Encode a bitmap-based ROA}
\label{sect:hroa:encode}

Given a set of authorized prefixes of an AS, a list of sub-tree blocks is produced with \texttt{ BM-ROA}, which are maintained with a hash table keyed by the sub-tree identifier. We call such a hash table a \emph{Sub-Tree Map} (STM in short). The encoding process is carried out prefix by prefix. 

For each authorized prefix $x/y$, where $x$ and $y$ are the IP address and the prefix length respectively, we can locate the sub-tree it belongs to and calculate its node number in this sub-tree in two steps. First, the prefix length $y$ is used to find out the nearest hanging level $l$ such that $l\leq y$.This process depends on the setting of hanging levels, which will be introduced in Section~\ref{sect:system:levels}. Then, the identifier of the sub-tree this prefix belongs to is made up by concatenating a set bit and the first $l$ bits of $x$. The node number of the node corresponding to this prefix, say $z$, is made up by concatenating a set bit and the middle $y-l$ bits of $x$ ended by the $y^{th}$ bit. Taking the authorized prefix \texttt{202.127.16.0/22} shown in Fig.~\ref{fig:example} as an example. Level-$20$ and level-$23$ are two hanging levels, between which level-$20$ is the nearest level the prefix can hang to. The identifier of this sub-tree it belongs to is calculated as 1878001, namely $\textbf{1}11001010011111110001$ in binary. Then, the middle $22-20=2$ bits of this prefix ended by the $22^{th}$ bit, namely $00$, are extracted and then appended to a set bit to make up node number 4 ($\textbf{1}00$ in binary).

\begin{figure}[tbp]
    \centering
    \includegraphics[width=\linewidth]{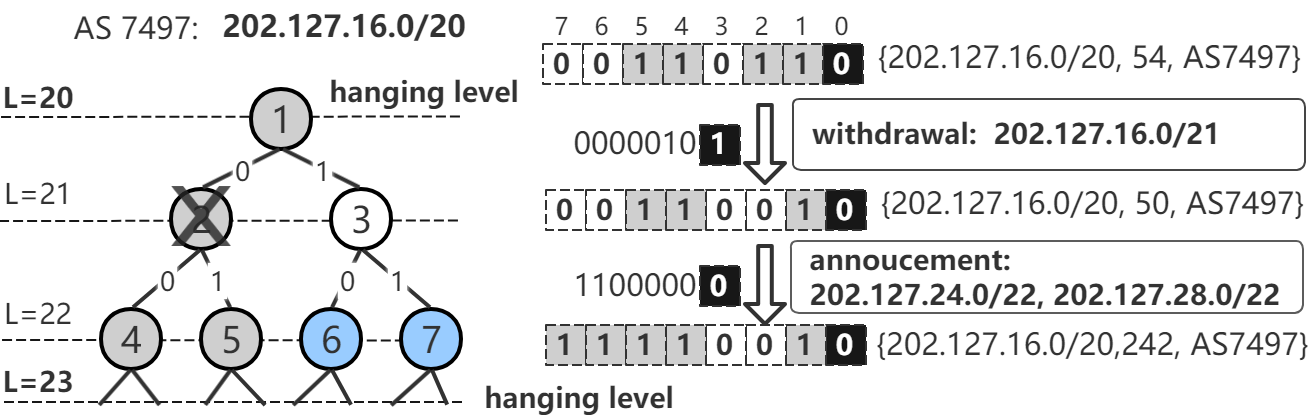}
    \caption{Process a batch of announcements / withdrawals with bitmaps.}
    \label{fig:update}
\end{figure}

Finally, an encoded sub-tree is obtained from the STM by the sub-tree identifier, or is initialized if the search of the sub-tree identifier fails. This sub-tree is then updated by setting the bit located by the node number $z$. Take the case shown in Fig.~\ref{fig:update} as an example. The two prefixes included in one authorization announcement target to the same sub-tree. Since their corresponding nodes are numbered as 6 and 7 respectively, the encoded sub-tree is updated with the corresponding two bits set.

Withdrawal of previous authorizations are processed in the same way, except that the withdrawal flag associated with every encoded sub-tree produced in the encoding process is set. As shown in Fig.~\ref{fig:update}, the withdrawal of the authorization of \texttt{202.127.16.0/21} produces an encoded sub-tree, where both bit-$0$ (the withdrawal flag) and bit-$2$ determined by the node number are set. Note that \texttt{BM-ROA} is able to process a batch of requests mixing with announcements and withdrawals, but they should be processed with two separate STMs.

Algorithm~\ref{alg:encoding} describes the process to encode a bitmap-based ROA with a batch of IPv4 prefixes. The process for IPv6 is similar.

\begin{algorithm}[t]
	\caption{Encode a hanging ROA with a batch of IPv4 prefixes}
	\label{alg:encoding}
    \SetKwInOut{Input}{Input}
    \Input{a batch of authorized prefixes of a given AS: $P$\\
and the withdrawal flag of this batch: $f$.
    }
	\KwOut{all sub-tree blocks of this hanging ROA: $B$.}
	\SetKwFunction{getLevel}{get\_nearest\_hanging\_level}
	\BlankLine
	
	Initiate an empty STM $M$ and and an empty list $B$.
	
	\ForEach{$(ip,~len)$ in $P$}{
	    $l \gets \getLevel(len)$\;
	    $w \gets (1 << (len - l)) - 1$\;
	    $x \gets  (1 << l) + (ip >> (32 - l))$\;
	    $y \gets  (1 << (len - l)) + ((ip >> (32 - len)) \And w)$\;
	    $M[x]\gets M[x] \mid (1 << y) $\;
	}

	\ForEach{(k,~v) in $M$}{
	    append the pair $(k,~v \mid f)$ to $B$\;
	}
\end{algorithm}

\subsection{Decode a bitmap-based ROA}
\label{sect:hroa:decode}
The key operation in decoding a bitmap-based ROA is to decode a batch of IP prefixes from a sub-tree block. This process takes two steps. First, from the sub-tree identifier we can decode the IP prefix $x/y$ represented by the root of this sub-tree. According to our encoding scheme, $y$ is actually the number of bits counted from the LSB to the most significant set bit of the sub-tree identifier. $x$ can be made up by shifting the $y$ bits ended by the LSB to the bit block starting from the MSB. 

Second, the position of every set bit (except the LSB) of the encoded sub-tree determines the node number of a node on this sub-tree. Then, its corresponding prefix can be constructed by expanding the prefix corresponding to the root of this sub-tree, with all bits of the node number except the MSB.

Algorithm~\ref{alg:decoding} describes the process to decode a batch of IPv4 prefixes from a sub-tree block of a bitmap-based ROA. The process for IPv6 is similar.

\begin{algorithm}[t]
	\caption{Decode a batch of IPv4 prefixes from a sub-tree block of a hanging ROA}
	\label{alg:decoding}

    \KwIn{the pair of an identifier and a bitmap: $(id, bm)$.}
    \KwOut{a list of IP prefixes: $P$;}
    \SetKwFunction{clz}{count\_leading\_zeros}
    \SetKwRepeat{Do}{do}{while}

	\BlankLine

	$l_1 \gets 31 - \clz (id)$\;
	$p_1 \gets (id \oplus (1 << l_1)) << (32 - l_1) $\;
	
	$bm \gets bm >> 1$\;
    \ForEach {setbit $i$ in $bm$}{
        $l_2 \gets 31 - \clz (i)$\;
        $p_2 \gets (i \oplus (1 << l_2)) << (32 - l_1 - l_2)$\;
    	append $(p_1 \mid p_2,~l_1 + l_2)$ to $P$\;
    }
\end{algorithm}

\subsection{Append a bitmap-based ROA}
\label{sect:hroa:append}
With \texttt{BM-ROA}, the cache of validated ROA payloads on the RP is maintained as a set of STMs, every two of them belong to an AS, and are used for announcements and withdrawals respectively. Given a bitmap-based ROA to append, we can use its ASN to locate two STMs. Then, every sub-tree block encoded in this bitmap-based ROA is inserted into the STM corresponding to its withdrawal flag. Besides, if the withdrawal flag of this sub-tree block is set, it should be inserted into the STM for announcements as well, so as to ensure the completeness of the cache response to a \texttt{reset} query in RTR. 

The key operation is to insert a sub-tree block into an STM. It takes two steps. First, an encoded sub-tree (say $x$) is obtained from the STM by the sub-tree identifier of the inserting block, or is initialized in case the search of the identifier in the STM fails. Then there are two cases to deal with. In the case that the encoded sub-tree (say $y$) contained in the inserting block has the same withdrawal flag as STM, $x$ is updated by $y$ with a bit-wise OR operation; otherwise, $x$ is updated by a bit-wise AND operation with $z$, which is produced by reverting all bits of $y$. Fig.~\ref{fig:update} demonstrates two examples of appending a sub-tree block with and without the withdrawal flag set respectively.

Algorithm~\ref{alg:append} describes the process of inserting a sub-tree block into an STM.

\begin{algorithm}[t]
	\caption{Insert a sub-tree block into an STM}
	\label{alg:append}

    \SetKwInOut{Input}{Input}
    \Input{the pair of an identifier and a bitmap: $(id, bm)$\\
           and an STM: $M$.
    }
    \SetKwFunction{bitwiseNot}{bitwise\_NOT}
    
	\BlankLine
	
    $f \gets bm \And 1$ \;
    \eIf{$M.flag = f$}{
        $M[id]\gets M[id] \mid bm$
    }{
        $M[id]\gets M[id] \And \bitwiseNot{bm}$
    }

\end{algorithm}

\subsection{Security, flexibility and scalability}
\label{sect:hroa:security}

% TODO: discuss these issues in intro
As for security, recent proposals~\cite{m-roa, m-roa-rfc} recommend to keep the \texttt{minimal ROA} principle, that only those IP prefixes that are actually authorized to originate from an AS can be attested with an ROA. Under the premise of adhering to this principle, scattered authorization cases, where a set of authorized IP prefixes of an AS cannot be maximally aggregated into a complete binary tree, are ubiquitous.
% 需要修改，根据ccdf图的结论
Accordingly, under the premise of implementing the \texttt{minimal ROA} principle, \texttt{BM-ROA} can provide same degree of security as that of \texttt{m-ROA} and offer better compression effects, that is, it generates fewer encoded entries.

Under the \texttt{minimal ROA} principle, route de-aggregation (e.g., for traffic engineering) requires explicit authorizations, which may lead to on-demand ROA updates. \texttt{BM-ROA} offers more flexibility in updating authorizations than \texttt{m-ROA}, by enabling batch processing with bitmaps. In addition, \texttt{BM-ROA} brings in great flexibility in ROA encoding with controllable compression. With proper setting of hanging levels, it can encode any part of the IP prefix trie into a sub-tree block, no matter whether its root represents an authorized prefix or not, and regardless of the number of authorized prefixes it covers. This offers even more flexibility in sub-tree encoding than \texttt{t-ROA}, and can lead to possible solutions for balancing availability and scalability when carrying multiple prefixes with an ROA~\cite{multi-prefix-roa}.

\texttt{BM-ROA} improves the scalability in two aspects. First, the encoding scheme offers an efficient and controllable manner to pack a set of authorized prefixes to compress ROAs. This can sharply reduce the number of ROA records, lightening the burdens of both transmission and processing. Second, with the hash tables keyed by the sub-tree identifier, the management of bitmap-based ROAs is truly scalable.

\section{The Hybrid ROA Encoding Scheme}
\label{sec:hybrid}
% 以图1中的例子，在面临small size block时，hanging roa能在保证安全的前提下提供较好的压缩效果。但是，无法处理huge size block，AS0 ROA中的authorized block属于这一类，极端情况下，一个高度为126的IPv6 block基于hanging roa编码将生成xxxxx条目，是无法接受的。
In cases of scattered authorization, \texttt{BM-ROA} outperforms \texttt{ML-ROA} (also \texttt{m-ROA} here) in terms of flexibility and scalability when keeping the \texttt{minimal ROA} principle. However, regarding to the current real-world RPKI system, there are always several CAs that do not follow the \texttt{minimal ROA} principle when authorizing IP prefixes, resulting in a set of concentrated authorization cases. Moreover, even if all CAs keep the \texttt{minimal ROA} principle, there will still exist some ROAs issued for special purposes, such as AS0 ROAs, whose origin ASN are set to AS0 and are used to “lock” unallocated
or unassigned IP address blocks; and ROAs used for DDoS mitigation and defensive de-aggregation in response to prefix hijacks. \texttt{BM-ROA} has limitations when it comes to handling the encoding of these special-purpose ROAs, especially AS0 ROAs because they are recommended to use with the \texttt{maxLength} set to 32 or 128 in IPv4 and IPv6, respectively, thus resulting in a set of clusters of highly concentrated  authorized IP prefixes. In the extreme case, an IPv6 address block with a height of 127 would necessitate the use of $2^{127}$ bits for encoding with \texttt{BM-ROA}, which is impractical, while \texttt{ML-ROA} tends to be the best encoding scheme, producing only one entry.
 % However, on one hand, during the transition phase, there will still be some CAs that do not follow the \texttt{minimal ROA} principle when authorizing IP prefixes, resulting in concentrated authorization cases in the real-world RPKI system. On the other hand, 
%On the other hand, if a CA prioritizes convenience over security, it may authorize an IP prefix and all its sub-prefixes in advance by specify maximum length without knowing its de-aggregation plan, to allow network operators to easily reconfigure their networks without modifying their ROAs. In this scenario, the IP prefixes that CAs authorize will fall under the category of concentrated authorization cases. 

In response to scenarios where scattered authorization cases and concentrated authorization cases coexist, we propose the hybrid ROA encoding scheme (\texttt{h-ROA} in short), which utilizes both \texttt{ML-ROA} and \texttt{BM-ROA} and encodes ROAs based on the pattern of authorized IP prefixes for a given AS, with the goal of maximizing compression efficiency.

\subsection{The trend and status quo of the height of address blocks in the real-world RPKI system}
\begin{figure}[tbp]
    \centering
    \includegraphics[width=\linewidth]{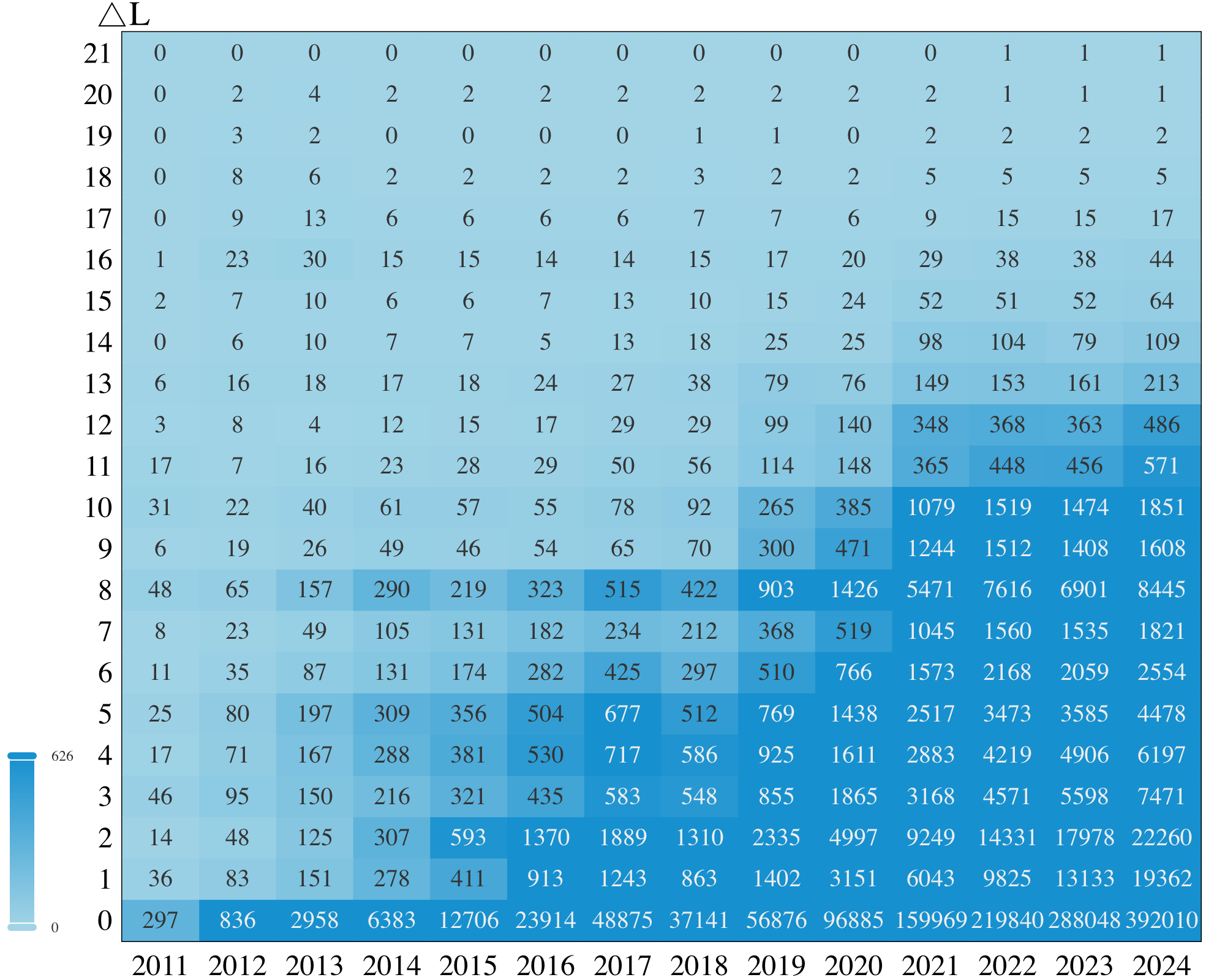}
    \caption{The trend of height of address blocks in IPv4.}
    \label{fig:height:v4}
\end{figure}

\begin{figure}[tbp]
    \centering
    \includegraphics[width=\linewidth]{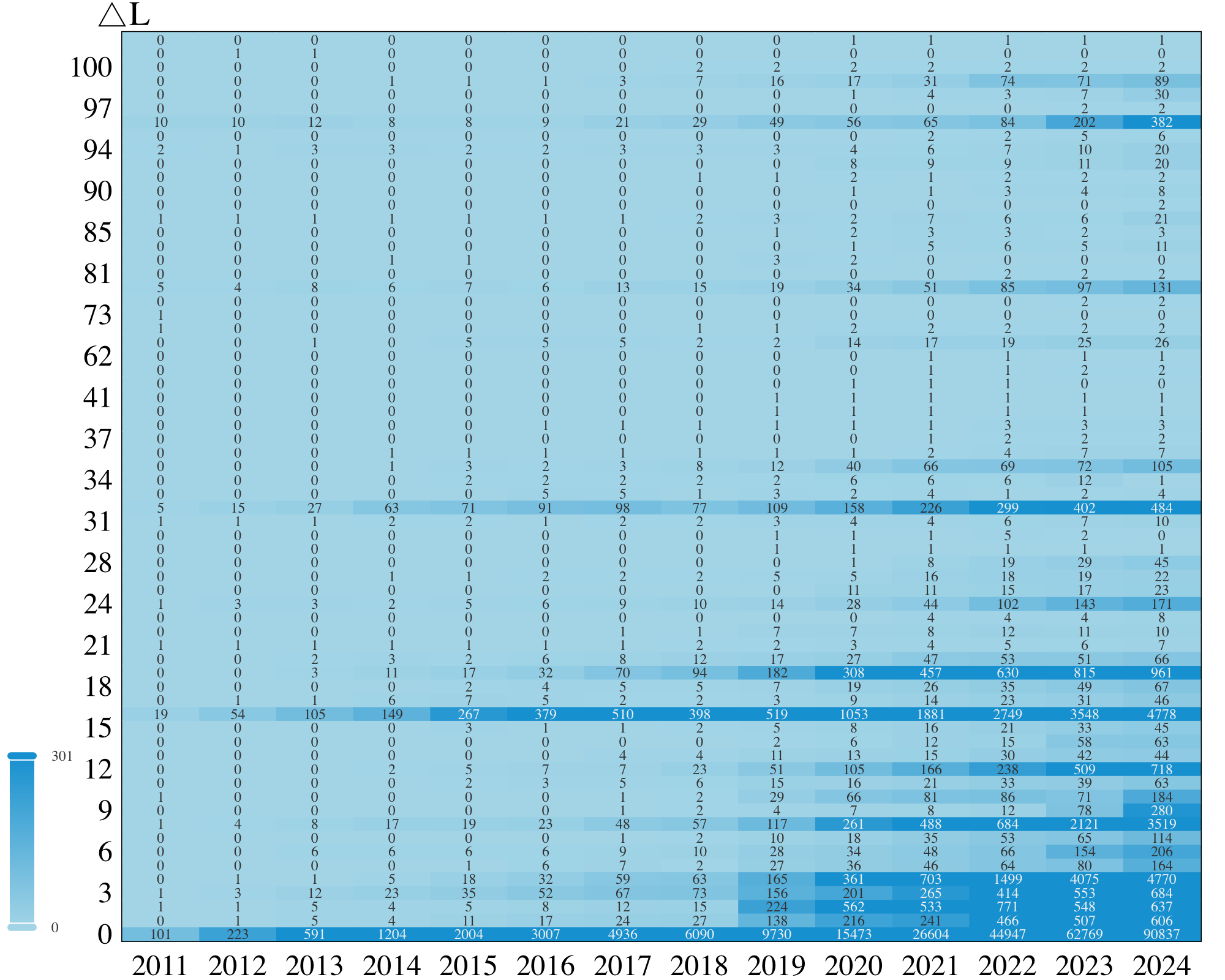}
    \caption{The trend of height of address blocks in IPv6.}
    \label{fig:height:v6}
\end{figure}

% block的size（△）的趋势（倾向于签发单一前缀）；block size的现状（仍然有huge size block的存在）
We define the ``height" of an address block, represented by the tuple (IP prefix, maximum length), as the difference between maximum length and the prefix length of this block, denoted as $\Delta{L}$, which influences the encoding efficiency of \texttt{BM-ROA}. Therefore, we extracted address blocks from the VRPs collected from the real-world RPKI system, for the time period from 1st Mar, 2011 to 1st Mar, 2024. On this basis, we did a statistical analysis on the height of every address block and the results are presented in Fig.~\ref{fig:height:v4} and~\ref{fig:height:v6}. In general, IPv6 address blocks tend to have a greater height compared to IPv4 address blocks, and there is a growing diversity in the distribution of height of  IPv6 address blocks, which is not reflected in IPv4 address blocks. Obviously, the AS0 address blocks with a height of 0, which means only one IP prefix is authorized, constitute the largest proportion every year for both IPv4 and IPv6. This trend indicates that most CAs are gradually becoming aware of the problems caused by \texttt{maxLength} and are choosing to deprecate it in favor of authorizing IP prefixes one by one. However, in recent years, IPv4 address blocks with a height exceeding 20 and IPv6 address blocks with a height exceeding 100 have continued to exist persistently, which will generate a large number of entries with \texttt{BM-ROA}.

% Given the current distribution of the height of address blocks, it may be unwise to encode a large-height address block with bitmaps. Therefore, it is crucial to develop an encoding strategy that determines which a set of authorized IP prefixes of an AS should be encoded using either \texttt{bitmap} or \texttt{maxLength}.

%
\subsection{How to determine whether an address block should be encoded with \texttt{BM-ROA} or \texttt{ML-ROA}?}
\label{sect:hybrid:delta}
In the current real-world RPKI system, both scattered authorization cases and large-height address blocks coexist, necessitating the development of a encoding strategy that combines the use of \texttt{ML-ROA} and \texttt{BM-ROA} to achieve the maximum compression effect. It is evident that encoding large-height address blocks with \texttt{BM-ROA} may not be efficient. Besides, the use of \texttt{ML-ROA} to encode extremely scattered authorized IP prefixes may render the compression capability of \texttt{maxLength} completely ineffective. This leads to a critical design question: how to determine whether an address block should be encoded using \texttt{bitmap} or \texttt{maxLength}?

To this end, we define the $\Delta{L}$ parameter as a criterion for selecting the appropriate ROA encoding scheme for a given address block between \texttt{ML-ROA} and \texttt{BM-ROA}. Specifically, if the $\Delta{L}$ of an address block is greater than or equal to the predefined threshold, then \texttt{ML-ROA} is employed for encoding; otherwise, \texttt{BM-ROA} is the chosen one. Furthermore, we select the optimal $\Delta{L}$ parameter by comparing the number and size of PDUs encoded with \texttt{h-ROA} under different $\Delta{L}$ parameter settings. More specifically, we construct three data sets, namely the minimal ROAs\footnote{If the minimal ROA satisfies the encoding conditions, meaning it can form several complete binary trees, we then encode them into address blocks with \texttt{maxLength}.} from the data set of \texttt{curr} and \texttt{full}, respectively, as well as the existing address blocks extracted from \texttt{verified ROAs} which consists of \textit{verified ROAs w/o AS0} and \textit{verified AS0} in Table~\ref{tb:data}. On this basis, we set different values to the $\Delta{L}$ parameter. For each $\Delta{L}$ parameter setting, we proceed to calculate the number and size of entries encoded using \texttt{h-ROA} across a range of hanging level configurations. Ultimately, we designate the $\Delta{L}$  parameter that yields the minimum number and size of PDUs as the optimal choice.

\begin{figure}[!tbp]
    \centering
    \includegraphics[width=\linewidth]{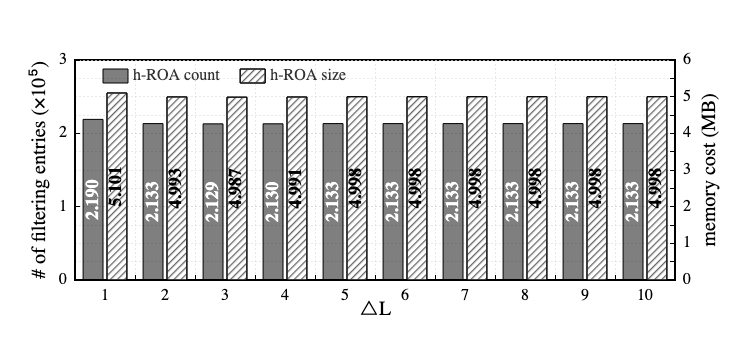}
    \caption{The total number and size of entries produced by encoding minimal ROAs of \texttt{curr} with different $\Delta{L}$ settings.}
    \label{fig:delta:rov}
\end{figure}

\begin{figure}[!tbp]
    \centering
    \includegraphics[width=\linewidth]{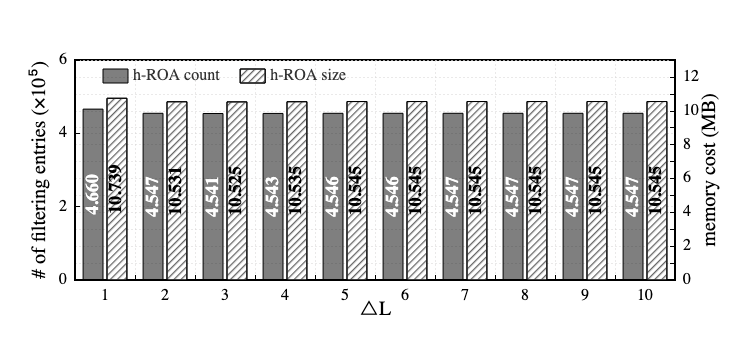}
    \caption{The total number and size of entries produced by encoding minimal ROAs of \texttt{full} with different $\Delta{L}$ settings.}
    \label{fig:delta:rib}
\end{figure}

\begin{figure}[!tbp]
    \centering
    \includegraphics[width=\linewidth]{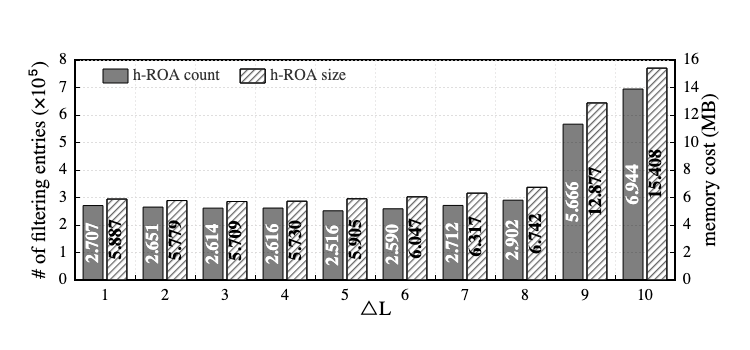}
    \caption{The total number and size of entries produced by encoding existing ROAs with different $\Delta{L}$ settings.}
    \label{fig:delta:vrp}
\end{figure}

Fig.~\ref{fig:delta:rov},~\ref{fig:delta:rib}, and~\ref{fig:delta:vrp} present the results, which indicate that setting the $\Delta{L}$ parameter to 3 achieves the most optimal results when encoding minimal ROAs from the \texttt{curr} and \texttt{full} data sets using \texttt{h-ROA}, in terms of both the total number and size of the encoded entries. While encoding existing ROAs from the \texttt{verified ROAs w/o AS0 ROAs} and \texttt{verified AS0 ROAs} data sets with \texttt{h-ROA} is optimal in terms of size, it does not necessarily achieve the best results in terms of number. After comprehensive consideration, we have decided to set the $\Delta{L}$ parameter to 3.

\subsection{How to determine the height of the sub-tree block to be used by \texttt{BM-ROA} for encoding?}
\label{sect:hybrid:level}
Given the authorized IP prefixes of \textit{all} ASes, the $\Delta{L}$ parameter is used to determine which AS's authorized IP prefixes are suitable for encoding with \texttt{ML-ROA} or \texttt{BM-ROA}. For those authorized IP prefixes designated for \texttt{BM-ROA} encoding, additional configuration is necessary, specifically regarding the \textit{hanging levels}. The setting of hanging levels dictates the height of sub-tree blocks used for encoding, which in turn affects the number and size of the resulting encoded entries. Like the previous Section~\ref{sect:hybrid:delta}, based on three datasets, namely the minimal ROAs from the data set of \texttt{curr} and \texttt{full}, respectively, as well as the existing address blocks extracted from \texttt{verified ROAs}, we evaluate the basic version of \texttt{h-ROA} where the hanging levels are set as multiples of a certain value. By adjusting this value and calculating the total size of the encoded entries for each hanging levels setting, with the $\Delta{L}$ parameter set to 3, we can obtain the optimal hanging level configuration.

\begin{figure}[!tbp]
    \centering
    \includegraphics[width=\linewidth]{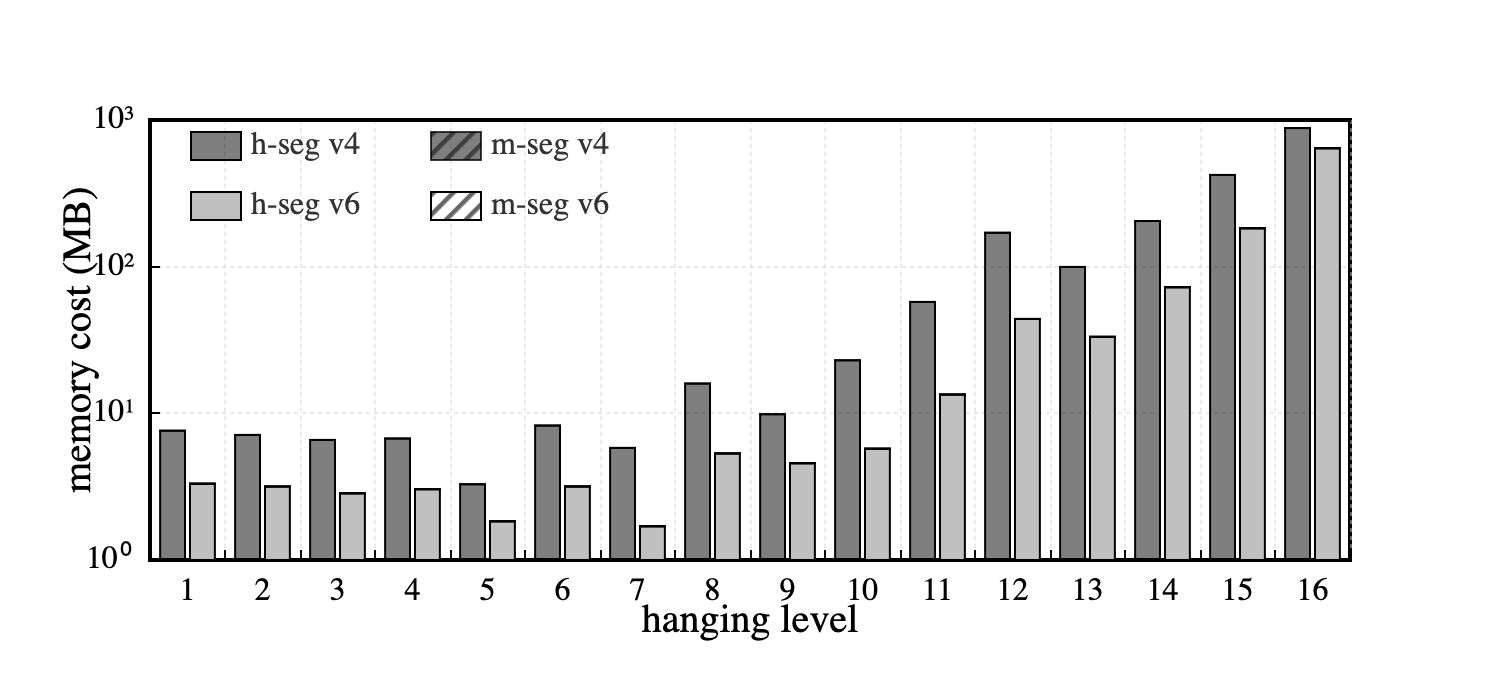}
    \caption{The total size of entries produced by encoding minimal ROAs of \texttt{curr} with different \textit{hanging level} settings.}
    \label{fig:level:rov}
\end{figure}

\begin{figure}[!tbp]
    \centering
    \includegraphics[width=\linewidth]{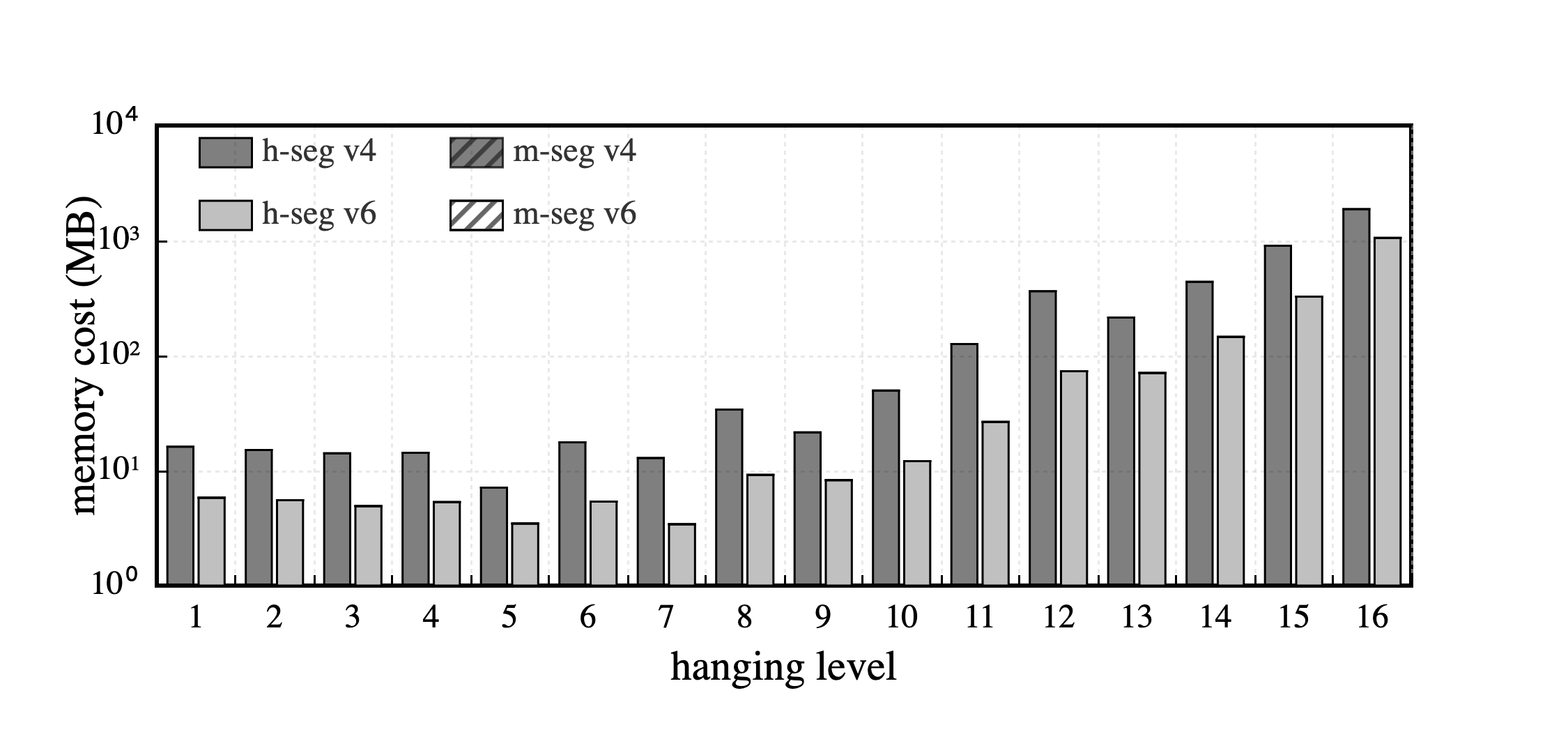}
    \caption{The total size of entries produced by encoding minimal ROAs of \texttt{full} with different \textit{hanging level} settings.}
    \label{fig:level:rib}
\end{figure}

\begin{figure}[!tbp]
    \centering
    \includegraphics[width=\linewidth]{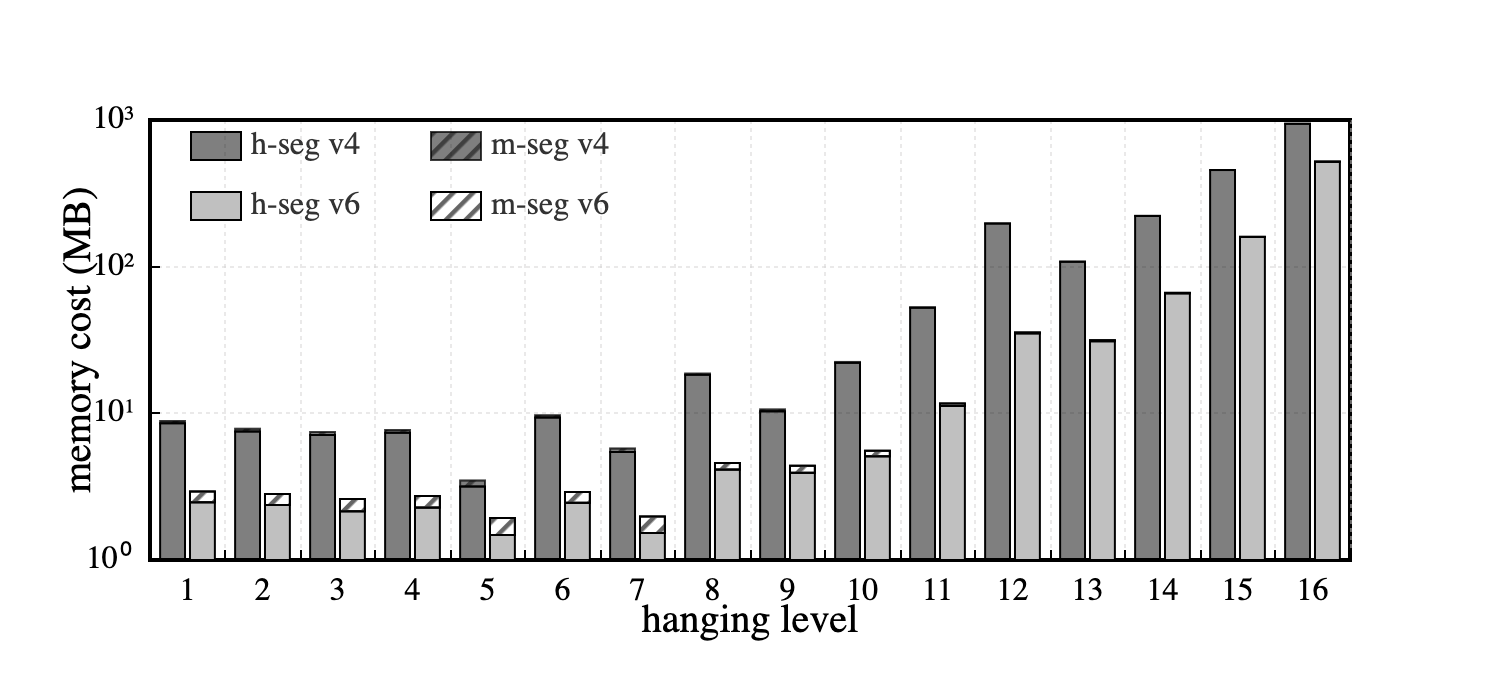}
    \caption{The total size of entries produced by encoding existing ROAs with different \textit{hanging level} settings.}
    \label{fig:level:vrp}
\end{figure}

Fig.~\ref{fig:level:rov},~\ref{fig:level:rib}, and~\ref{fig:level:vrp} present the results. Given that the $\Delta{L}$ parameter is set to 3, the configuration where every hanging level is a multiple of 5 achieves the best compression effect across all datasets, with the smallest total size of encoded entries. In the next system implementation (Section~\ref{sect:system}) and experimental evaluation (Section~\ref{sect:exp}), we configure the $\Delta{L}$ parameter and \textit{hanging level} parameter to 3 and 5, respectively. 

\section{System Implementation}%1.5p
\label{sect:system}

\subsection{Overview of the prototype implementation}
% discuss different kinds of deployment and the modifications needed. CA+RP+Router, RP+Router
% introduce the overview of the reference implementation

As shown in Table~\ref{tb:deployment}, the implementation of \texttt{h-ROA} is flexible and involves a series of functions on different entities of RPKI. Regarding a full implementation, the CA software needs an encoder to encode hanging ROAs, the RP should maintain validated ROA payloads with the function of appending hanging ROAs, and the router has to refactor its route validator to use hanging ROAs\footnote{the design of such a validator is beyond the scope of this work.}. To minimize the modifications to existing protocols and software packages, we take the minimal implementation, where only the RP and the router require some minor changes. On the two sides, open-source software packages \texttt{RPstir2}\footnote{RPstir2: [https://github.com/bgpsecurity/rpstir2]} and \texttt{Routinator}\footnote{Routinator:[https://github.com/NLnetLabs/routinator10]} are adopted as RPs, and \texttt{goBGP}\footnote{goBGP:[https://github.com/osrg/gobgp]} and \texttt{FRRouting}\footnote{FRRouting:[https://github.com/FRRouting/frr]} are employed as routers for the prototype implementation, where only the \texttt{reset} query is processed with \texttt{h-ROA}.

\begin{table}[!tbp]
\caption{Functionalities required for \texttt{h-ROA} implementation}
\begin{center}
 \begin{tabular}{|l|c|c|c|}\hline
        & CA                     & RP                    & Router                  \\ \hline
Full    & \texttt{hroa\_encode}  & \texttt{hroa\_append} & \texttt{hroa\_validate} \\\hline
Minimal &                        & \texttt{hroa\_encode} & \texttt{hroa\_decode}   \\\hline
\end{tabular}
\label{tb:deployment}
\end{center}
\end{table}

\subsection{PDU format and configuration of hanging levels}
\label{sect:system:levels}

According to the tuning results of the \textit{hanging level} parameter in Section~\ref{sect:hybrid:level}, we adopt a configuration where every hanging level is a multiple of 5 and the distance between any two successive hanging levels is exactly 5, which ensures that an encoded sub-tree consumes at most 32 bits. Besides, the depth of the last hanging level will not exceed $31$ or $127$ in IPv4 and IPv6, and this feature ensures the sub-tree identifier is confined within 32-bits or 128-bits bound for IPv4 and IPv6, respectively. 

\begin{figure}[!tbp]
  \begin{minipage}[b]{0.35\linewidth}
    \centering \includegraphics[width=0.7\textwidth]{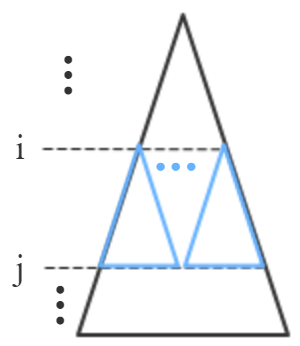}
    \caption{Hanging levels.}\label{fig:level}
  \end{minipage}%
  \hfill
  \begin{minipage}[b]{0.65\linewidth}
    \centering
    \includegraphics[width=\textwidth]{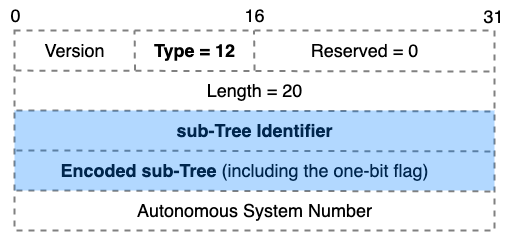}
    \caption{RTR IPv4 prefix PDU.}\label{fig:pdu}
  \end{minipage}
\end{figure}

We define two new types of RTR payload PDUs for IPv4 and IPv6 respectively. Fig.~\ref{fig:pdu} shows the new IPv4 prefix PDU, which can reuse fields from the original PDU and ensure that its size remains same as the original one. A new type (12) is assigned to it in order to support hybrid deployment mixing with \texttt{h-ROA} and existing schemes. In addition, the middle 64-bit block is refactored to encode a sub-tree and its identifier, each taking 32 bits. The new IPv6 prefix PDU, with the type set to 13, has the same format except that the sub-tree identifier takes 128 bits. 

% This PDU format brings in two restrictions in setting hanging levels. First, the distances between successive hanging levels, which determine the height of sub-trees, must not exceed $5$.  This ensures that an encoded sub-tree consumes at most 32 bits. Besides, to keep the sub-tree identifier within its bound, the depth of the last hanging level must not exceed $31$ or $127$ in IPv4 and IPv6 respectively. As such, we adopt a straightforward configuration where every hanging level is a multiple of 5 and the distance between any two successive hanging levels is exactly 5.

%At present, there are two types of deployment. The first uses CA, RP and router(deployment A), and the second only uses RP and router(deployment B). The system modifications required for the two deployments are shown in table. For deployment A, CA needs to aggregate and encode prefixes before sending packet and append it in RP. After receiving RTR packet from RP, router decodes and verifies prefixes. The whole process involves three supporting objects and causes more modification on servers. As a result, we select deployment B, in which RP not only encodes the prefixes, but also append ROA, and router does the same as Deployment A. Through the above deployment, we have achieved the minimal modification to the system. 

%In order to minimize the modification, we only use 64 bits to save sub-tree identifier and encoded sub-tree, which acquires the selection of changing level not exceeding 31. Besides, If hanging level is less than 5, the complement of zero in the alignment process will cause too much waste. Further, in the follow-up experiments, we found that most IPv4 prefixes are clustered in the 24th layer of sub tree. As a result, We select the multiples of 5 as the hanging levels when encoding sub-tree. On the one hand, this selection can ensure that the generated PDU size will not be larger than the PDU size specified in the previous protocol. On the other hand, it can ensure that a PDU contains at least one encoded prefix corresponding to a bitmap. The advantage of this selection is to ensure that the ROA payload of a PDU is not less than that of the original PDU.

\subsection{Encode hybrid ROAs on the RP}
\label{sect:system:encode}

In response to the \texttt{reset} query, the RP loads all verified ROAs, each containing an ASN and a set of prefixes, from the local storage, and encodes those prefixes need to be encoded with \texttt{BM-ROA} into a set of bitmap-based ROAs using the method described in Algorithm~\ref{alg:encoding}. Bitmap-based ROAs are maintained as sub-tree blocks using a set of STMs. Every two of them belong to an AS, for the convenience of announcement and withdrawal respectively. The maintenance for maxLength-based ROAs remains unchanged, depending on the specific implementation of the RP software. Finally, for every AS involved, all sub-tree blocks stored with its STM and address blocks for announcements are encapsulated into a list of RTR payload PDUs, one for each.

\subsection{Decode hanging ROAs on the router}
\label{sect:system:decode}

In receiving a PDU response from the RP, the router checks its PDU type to determine whether to process it with \texttt{BM-ROA}. In case it encapsulates a bitmap-based ROA, a sub-tree block and the ASN can be decapsulated from it. Further, the router decodes a set of IP prefixes from this sub-tree block using the method described in Algorithm~\ref{alg:decoding}, which are then passed on to the route validator as normal.

% \subsection{Deal with AS0 ROAs}
% \label{sect:system:as0} AS0 ROAs, whose origin ASN are set to AS0, are used to ``lock" unallocated or unassigned IP address blocks, and are recommended to use with the \texttt{maxLength} set to 32 or 128 in IPv4 and IPv6 respectively. They have a lower priority than other ROAs in affecting the route's validity state~\cite{rfc6483-rov}, and are not widely used in current RPKI systems. As such, we do not compress AS0 ROAs, and process them as normal.

%PDU is decoded into an ASN as well as a set of (\texttt{IP}, \texttt{length}, \texttt{maxLength}) tuples in . router, the format of the tuples is the same as that in the original router. The obtained tuples are inserted into the data structure maintained by the router (which can be tables or trees), then tuples are used to verify the received (\texttt{asn}, \texttt{prefix}) key-value pairs. The above process only needs to add a decoding module to the router after it receives the RTR packet and decapsulates the packet into corresponding PDU, without modifying the data structure maintained by router, nor will it have any impact on the subsequent verification process in router. In system implementation, we use the open source \texttt{goBGP} as software router.

%An AS0 ROA may coexist with ROAs that have different subject AS values; although in such cases, the presence or lack of presence of the AS0 ROA does not alter the route's validity state in any way\cite{as0-rfc}.
    
\section{Further compression}%1p
\label{sect:opt}

In this section, we discuss in depth the techniques we use to achieve further compression with \texttt{BM-ROA} in two aspects. %\emph{optimizing the setting of hanging levels} and \emph{aggregating PDUs sharing the same origin ASN}. The former aims at finding an optimal way to group a set of authorized prefixes of an AS into sub-tree blocks, while the latter is to aggregate and assemble them as a whole into one PDU.   

\subsection{Optimize the setting of hanging levels}
\label{sect:opt:dp}

This technique aims at finding out an optimal way to group a set of authorized prefixes of an AS into sub-tree blocks. By representing these prefixes with the IP address trie, we define $opt(i)$ as the minimized total size of PDUs required to encode all authorized prefixes in the first $i$ levels (numbered from 0) of the IP address trie. Suppose level-$i$ and level-$j$ are set as two successive hanging levels, the total size of PDUs required to encode all authorized prefixes between level-$i$ and level-$j$ (not included) can be calculated as $cost(i,j)= num \times size$, where $num$ is the number of non-empty sub-trees rooted at level-$i$ and ending by level-$j$ (not included), while $size$ is determined by the size of the sub-tree identifier and the length of the bitmap ($2^{j-i}$). Then, we have $opt(j)= \min^{j-1}_{i=1} \{opt(i) + cost(i,j)\}$, and the solution to $opt(i)$ can be found with dynamic programming.

The optimal solution may come with an uneven distribution of hanging levels. As a result, sub-trees rooted at different hanging levels may have different heights, requiring variable-length bitmaps to encode them.

\subsection{Aggregate PDUs sharing the same origin ASN}
\label{sect:opt:aggregate}

Given a set of prefix PDUs of the same type that share the same origin ASN, there is a lot of redundant information among them. Take two IPv4 prefix PDUs encoded with \texttt{BM-ROA} as an example. Since they share the same origin ASN, only the sub-tree identifier and the encoded sub-tree are identical for each of them, which takes only 8 bytes out of the total 20 bytes. Therefore, packing them together into one PDU with redundant information eliminated can reduce the number as well as the total size of PDUs. 

To aggregate a set of PDUs with sub-tree blocks encoded into fixed-length bitmaps, all these sub-tree blocks are put together, one after another. Since each sub-tree block takes 8 bytes, the number of sub-tree blocks contained in a PDU can be implied by its length. Besides, with the first sub-tree block being at a fixed position inside the PDU, every other one can be easily located. However, the aggregation of PDUs with variable-length bitmaps is more complex. Every sub-tree block to be packed together is associated with a \texttt{height} field that indicates the height of the sub-tree encoded in this block. 
\section{Experimental Evaluation}%2.5p
\label{sect:exp}
In this section, we evaluate the performance of \texttt{h-ROA} with extensive experiments, and verify its superiority in comparison with  \texttt{t-ROA}~\cite{rfc6482-roa} and  \texttt{m-ROA}~\cite{m-roa}. \texttt{t-ROA} is the standardized solution widely used in current RPKI systems, and \texttt{m-ROA} is the state-of-the-art approach and under the standardization~\cite{m-roa-rfc}.  
 
\subsection{Methodology}% 0.25p
% datasets: current AND future; v4 AND v6
% approaches: t-roa (best), m-roa, h-roa
% algorithm evaluation VS. system evaluation
\subsubsection{Data sets}
\label{sect:exp:data}
We collected data from real-world RPKI and BGP systems, and put them into two data sets referred to as \emph{curr} and \emph{full} respectively. Using the RP software \texttt{Routinator} pre-configured with the trust anchor locators of all the five regional Internet registries, we collected all ROAs from the global RPKI repository on March 1st, 2024 and validated them following RPKI polices. In this way, we obtained two sets of ROAs, the received ROAs and the verified ROAs, excluding all AS0 ROAs that are maintained separately. Then we collected 1,214,817 unique RIB entries from public route collector RRC00 from the Routing Information Service (RIS)~\cite{ripe-ris} on the same day. We extracted the (prefix, origin ASN)-mappings from those RIB entries that are covered by received ROAs and verified ROAs respectively, and constructed two sets of minimal ROAs as shown in Table~\ref{tb:data}. They along with the collected ROAs comprise the \emph{curr} data set. Regarding the full deployment of RPKI in the near future~\cite{m-roa}, all RIB entries would be covered by ROAs. Accordingly, all the (prefix, origin ASN)-mappings extracted from collected RIB entries form a set of minimal ROAs, which is the \emph{full} data set. In the same way, we collected 21 RIBs from the route collector \emph{RRC00}\footnote{\url{https://data.ris.ripe.net/rrc00/}} from 2004 and 2024, one per year on the same date March 1st, and constructed 21 sets of minimal ROAs for scalability evaluation.

\begin{table}[!tbp]
\centering
\caption{Data set specification}\label{tb:data}
\begin{threeparttable}
\begin{tabular}{|c|c|c|c|c|c|}
\hline
\multirow{2}*{}     & \multicolumn{3}{c|}{\# of collected ROAs / RIB entries\tnote{1}}    & \multicolumn{2}{c|}{\# of minimal ROAs\tnote{2}} \\ \cline{2-6} 
                      &                               & IPV4         & IPV6        & IPV4               & IPV6               \\ \hline
full                  & RIB entries             &1003439       &211378            & 1003439             & 211378            \\ \hline
\multirow{4}*{curr} & received w/o AS0           & \multicolumn{2}{c|}{237208} &  477245                  & 121118                   \\ \cline{2-6} 
                      & verified w/o AS0            & \multicolumn{2}{c|}{196517}      & 473950             & 120391              \\ \cline{2-6} 
                      & received AS0       & \multicolumn{2}{c|}{596}   & N/A                  &N/A                   \\ \cline{2-6} 
                      & verified AS0       & \multicolumn{2}{c|}{537}      & N/A               & N/A              \\ \hline
\end{tabular}
\begin{tablenotes}
\item[1] collected from real-world RPKI and BGP systems on January 1st, 2022.
\item[2] extracted from the RIB or the intersection of the RIB and collected ROAs.
\end{tablenotes}
\end{threeparttable}
\end{table}

\subsubsection{Evaluated schemes}
For \texttt{h-ROA}, we evaluate its basic version where the hanging levels are set as multiples of 5 (introduced in Section~\ref{sect:system}), along with its extension with PDU aggregation (introduced in Section~\ref{sect:opt}), which are referred to as the \emph{h-ROA}  and the \emph{ah-ROA} respectively.
For \texttt{m-ROA}, we use its official implementation from the open-source project\footnote{https://github.com/yossigi/compress\_roas} for performance evaluation, and re-implement it into \texttt{GoBGP} and \texttt{FRRouting} with GO and C languages, respectively, for system integration.

% along with its three extensions (introduced in Section~\ref{sect:opt}). The basic version and its extension with PDU aggregation are referred to as the \emph{fixed-length-bitmap} (FLB) and the \emph{aggregated FLB} (AFLB) respectively. The approach of optimizing the setting of hanging levels with a dynamic programming algorithm is referred to as \emph{variable-length-bitmap} (VLB). Its extension with PDU aggregation is referred to as \emph{aggregated VLB} (AVLB). In the implementation of \texttt{t-ROA}, we adopt the maximally-permissive strategy~\cite{m-roa} to maximize the compression by the use of \texttt{maxLength}. 

\subsubsection{System implementation}
For system evaluation, we construct a small RPKI system with four open sourced software packages, \texttt{rpstir2}, \texttt{Routinator}, \texttt{goBGP} and \texttt{FRRouting}, which act as the RPs and the routers respectively. In our prototype implementation (as introduced in Section~\ref{sect:system}), the basic version of \texttt{h-ROA} is adopted, where the encoding and decoding functions are integrated into the corresponding modules of \texttt{rpstir2}, \texttt{Routinator}, \texttt{goBGP} and \texttt{FRRouting} respectively. For \texttt{m-ROA}, we implement its compression algorithm with an encoding function integrated into \texttt{rpstir2} and \texttt{Routinator} as well. 

% There are no changes on the two ends for \texttt{t-ROA}.

\subsubsection{Running platforms}
We run all system experiments on two cloud servers from different providers. We refer to these two servers as \emph{cloud-1} and \emph{cloud-2} respectively. The specifications of them can be found in Table~\ref{tb:server}. Besides, we evaluate the encoding and decoding speed of ROA schemes on three platforms with different CPUs. The specifications of them can be found in Table~\ref{tb:cpu}.
%TODO: 补充三个CPU的信息

\begin{table}[!tbp]
\centering
\caption{Cloud Specification}\label{tb:server}

\begin{tabular}{|c|c|c|}\hline
\multirow{3}*{cloud-1} 
& CPU  & Intel Xeon Platinum 8269CY (2.5GHz, 8Cores) \\ \cline{2-3} 
& RAM  & DIMM 16GB                                   \\ \cline{2-3}
& Provider & \url{https://www.aliyun.com/} \\ \hline
\multirow{3}*{cloud-2} 
& CPU  & AMD EPYC 7742 (2.2GHz, 8Cores)               \\ \cline{2-3} 
& RAM  & DIMM 16GB                                    \\ \cline{2-3}
& Provider & \url{http://www.cstcloud.cn/} \\ \hline
\end{tabular}
\end{table}

\begin{table}[]
\caption{CPU Specification}\label{tb:cpu}
\resizebox{\columnwidth}{!}{%
\begin{tabular}{|l|l|l|l|}
\hline
Model        & Xeon Platinum 8160 & EPYC 7742 & Kunpeng920 \\ \hline
Architecture & x86\_64            & x86\_64   & ARM        \\ \hline
Provider     & Intel              & AMD       & Huawei     \\ \hline
\end{tabular}%
}
\label{tab:my-table}
\end{table}

\subsubsection{Performance metrics}
Given a mapping between ASes and the prefixes they are authorized to originate, the 
\emph{number of PDUs} produced with every scheme is measured to evaluate the compression effect of ROA encoding. 
%Further, the \emph{compression ratio}, that indicates the degree of the compression an encoding scheme achieves, is measured as the average number of authorized prefixes one PDU carries. 
The \emph{total size of ROA payloads} in RRDP and the \emph{total size of payload PDUs} in RTR are measured to represent the transmission cost of ROAs and PDUs. In addition, the \emph{encoding} / \emph{decoding speed} is measured as the average number of authorized prefixes the encoding / decoding function reads / outputs per second, in MPPS (million prefixes per second). In the system evaluation, the time elapsed between the time when the router sends an RTR \texttt{reset} query and that when it receives the \texttt{End of Data} PDU is measured as the \emph{synchronization time}. We run every experiment 100 times and report the average results.

 E\subsection{Compression effect}
\label{exp:compress}
We evaluate the compression effects in encoding ROAs with two schemes, %\texttt{t-ROA}, 
\texttt{m-ROA} and \texttt{h-ROA}, using the two data sets \texttt{curr} and \texttt{full} that correspond to the current practice of RPKI and its full deployment in the future respectively. Two encoding formats, the ROA payload encapsulated in an RRDP packet as well as the RTR payload PDU, are considered, which are referred to as \emph{ROA} and \emph{PDU} respectively in short.

Of the data set \texttt{curr}, the received ROAs are exactly the ROAs encoded with \texttt{t-ROA}, and used as the ROAs encoded with \texttt{m-ROA} and as the input to produce hybrid ROAs with \texttt{h-ROA}.
%while the PDUs are directly constructed from the verified ROAs. The minimal ROAs corresponding to the received ROAs are used as the ROAs encoded with \texttt{m-ROA}, and as the input to produce hanging ROAs with \texttt{h-ROA}. At last, 
 The minimal ROAs corresponding to the verified ROAs and the PDUs are directly constructed from the verified ROAs are used as the input to produce PDUs with \texttt{m-ROA}, \texttt{h-ROA} and its extension with PDU aggregation \texttt{ah-ROA}, respectively. Of the data set \texttt{full}, the minimal ROAs corresponding to the whole RIB are used as the input to produce PDUs with \texttt{m-ROA}, \texttt{h-ROA} and its extension with PDU aggregation, respectively.
%the minimal ROAs corresponding to the whole RIB are used as the ROAs encoded with \texttt{m-ROA}, and as the input to produce hanging ROAs and maximally-permissive ROAs with \texttt{h-ROA} and \texttt{t-ROA} respectively. For \texttt{t-ROA}, the PDUs are constructed from the maximally-permissive ROAs directly. As for \texttt{m-ROA} and \texttt{h-ROA}, the PDUs are constructed from the minimal ROAs corresponding to the whole RIB. It is noteworthy that the maximally-permissive ROAs reflect nearly the best compression effect that can be achieved by the use of \texttt{maxLength}, but should never be deployed in reality due to its extreme lack of security~\cite{m-roa}.

\begin{figure*}[!tbp]
    \centering
    \includegraphics[width=\textwidth]{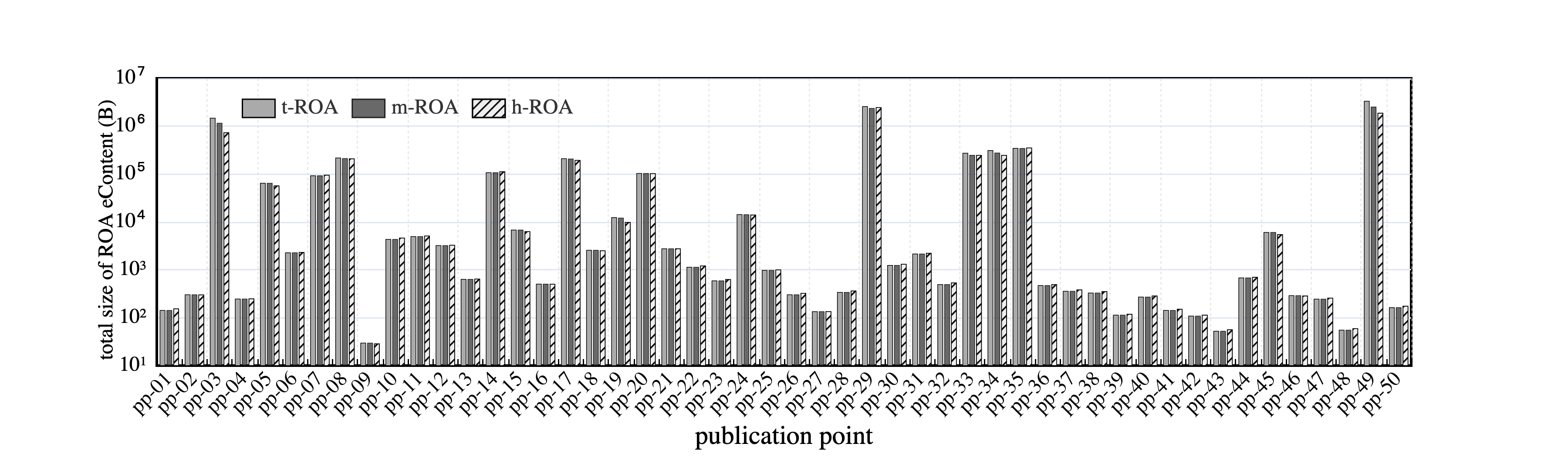}
    \caption{Transmission cost in conveying ROAs from the CA to the RP.}
    \label{fig:rrdp:cost}
\end{figure*}

The Fig.~\ref{fig:rrdp:cost} shows the compression
effects achieved with the \texttt{t-ROA}, \texttt{m-ROA} and \texttt{h-ROA}, in terms of transmission cost in conveying ROAs from the CA to the RP, which are measured as the the total size of ROA payloads encapsulated in RRDP packets. In most cases, \texttt{h-ROA} has a higher compression ratio than the other two schemes, and can compress the total size of
ROA payloads in RRDP by 23.4$\%$ and 26.6$\%$ compared with \texttt{t-ROA} and \texttt{m-ROA} respectively. However, in few cases, the total size of ROA payloads encoded with \texttt{h-ROA} is larger than that of \texttt{t-ROA} and \texttt{m-ROA}. The main reason is that these ROAs contain only one authorized IP prefix, and the deprecation of \texttt{maxLength} leads to a smaller payload size.
%TODO: 23.4和26.6

\begin{figure}[!tbp] 
  \begin{minipage}[b]{0.48\linewidth}
    \centering
    \includegraphics[width=\textwidth]{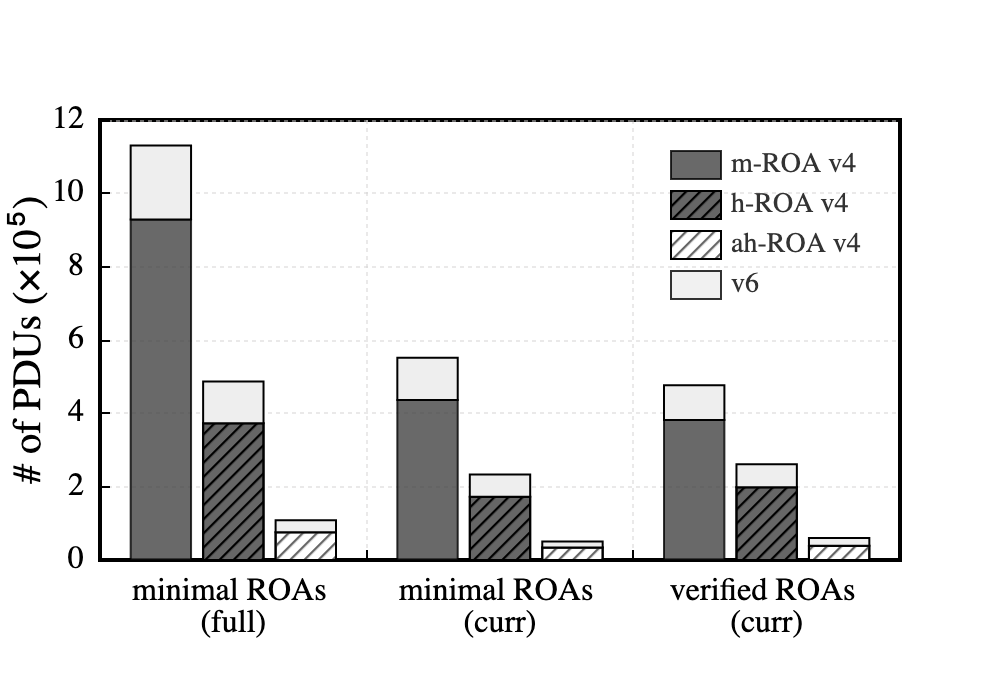}
    \caption{The number of PDUs encoded with three schemes.}\label{exp:rtr:count}
  \end{minipage}%
  \hfill
  \begin{minipage}[b]{0.48\linewidth}
    \centering
    \includegraphics[width=\textwidth]{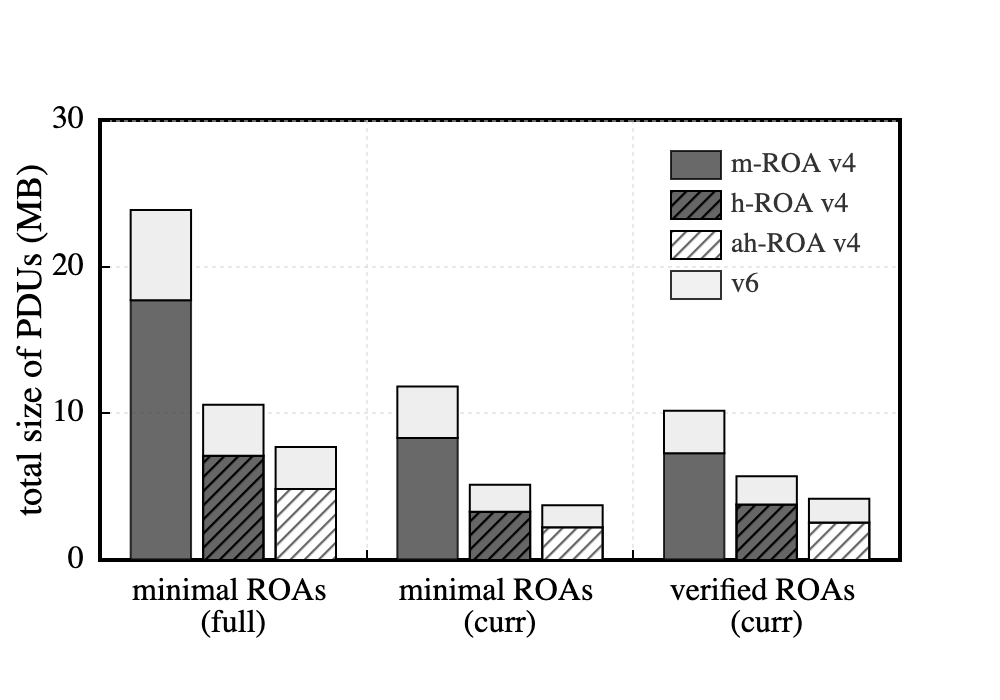}
    \caption{The size of PDUs encoded with three schemes.}\label{exp:rtr:size}
  \end{minipage}
\end{figure}

The Fig.~\ref{exp:rtr:count} and~\ref{exp:rtr:size} shows the compression effects achieved with \texttt{m-ROA}, \texttt{h-ROA} and its extension with PDU aggregation \texttt{ah-ROA}, in terms of the number of PDUs and total size of them, respectively. Clearly, \texttt{h-ROA} has a much higher compression ratio than \texttt{m-ROA}, where \texttt{h-ROA} reduces the number and size of PDUs by more than 45.1$\%$ and 43.9$\%$ in comparison with \texttt{m-ROA}. 
% h-roa和m-roa比较，number有3个值，最小值对应X，size有3个值，最小值对应Y
Moreover, by aggregating multiple PDUs sharing the same origin ASN into one, \texttt{ah-ROA} can achieve further compression, the number of PDUs can be sharply reduced, while the compression on
their sizes is not that obvious. More specifically, compared with \texttt{h-ROA}, the number and total size of PDUs can be reduced by 77.6$\%$ and 27.2$\%$ respectively with minimal ROAs of data set \texttt{full}, which turns to be 78.4$\%$
%ah-roa和h-roa比；
and 27.1$\%$ with minimal ROAs, and 77$\%$ and 26.9$\%$ with verified ROAs of data set \texttt{curr}.These two metrics together determine the overhead of the router to fetch validated ROAs from the RP.

\subsection{Scalability}
\label{exp:scalability}

\begin{figure}[!tbp] 
  \begin{minipage}[b]{0.48\linewidth}
    \centering
    \includegraphics[width=\textwidth]{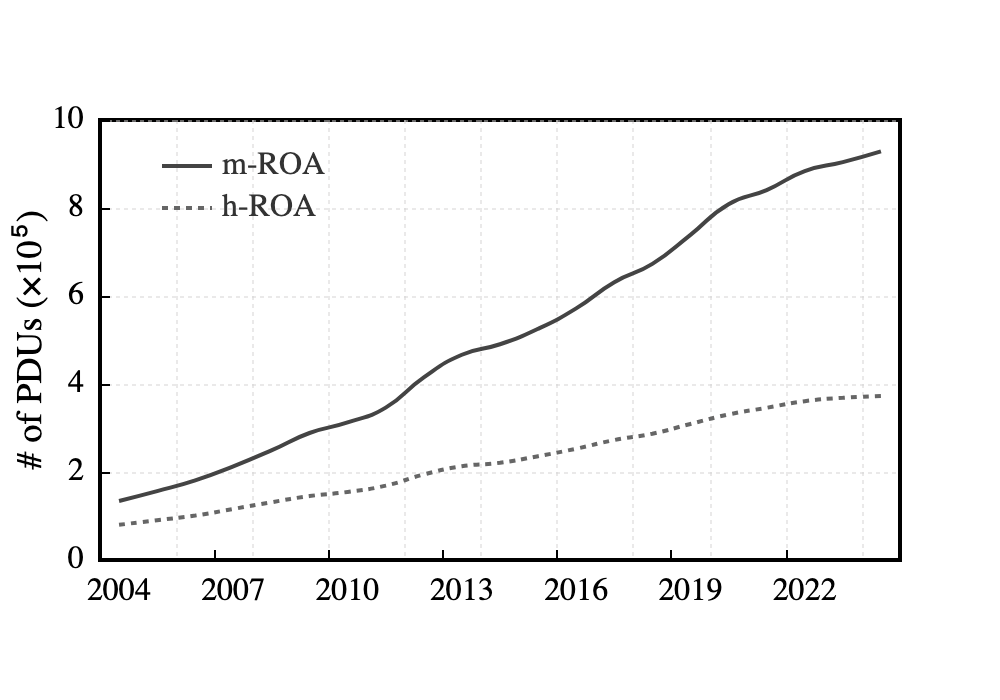}
    \caption{Scalability in IPv4.}\label{exp:scalability:v4}
  \end{minipage}%
  \hfill
  \begin{minipage}[b]{0.48\linewidth}
    \centering
    \includegraphics[width=\textwidth]{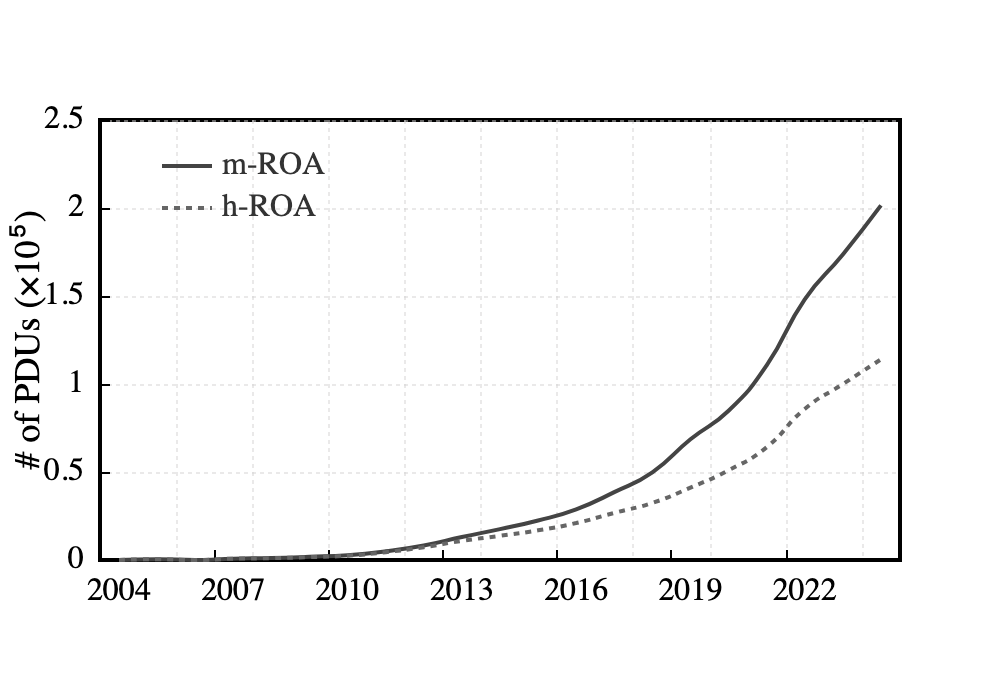}
    \caption{Scalability in IPv6.}\label{exp:scalability:v6}
  \end{minipage}
\end{figure}

We evaluate the scalability of each scheme via the increase of the number of PDUs produced over 21 sets of minimal ROAs, each of which is constructed from RIB entries collected at the same collector on the same date of one year. As shown in Fig.~\ref{exp:scalability:v4}, the scalability of \texttt{h-ROA} is much higher than \texttt{m-ROA}. More specifically, with the increase of the number of minimal ROAs to encode over 20 years, the number of IPv4 PDUs produced with \texttt{h-ROA} increases by $3.66$ times, while the ones produced with \texttt{m-ROA} increase by $5.91$ times. In IPv4, the superiority of \texttt{h-ROA} over \texttt{m-ROA} remains clear, and this gap continues to widen. From the trend demonstrated in Fig.~\ref{exp:scalability:v6}, we can deduce that the superiority of \texttt{h-ROA} over \texttt{m-ROA} will become clearer with the increase of the number of IPv6 routes.
% This is because the IPv6 routes used today are too few compared with the huge address space available. 
%TODO:X和Y，2024年的减去2004年的/2004

\subsection{Encoding / Decoding Speed}
\begin{figure}[!tbp]
    \centering
    \includegraphics[width=\linewidth]{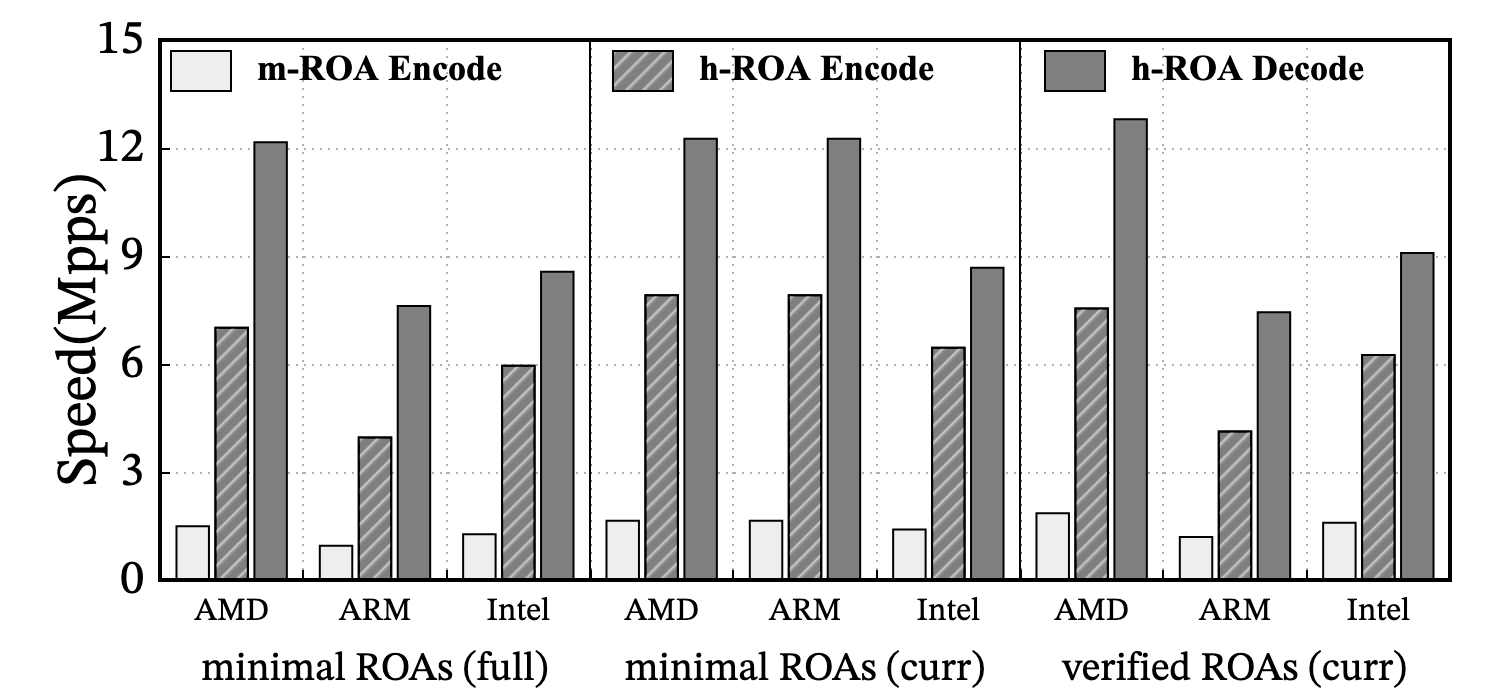}
    \caption{Encoding / Decoding speed in IPv4.}
    \label{fig:speed:v4}
\end{figure}

\begin{figure}[!tbp]
    \centering
    \includegraphics[width=\linewidth]{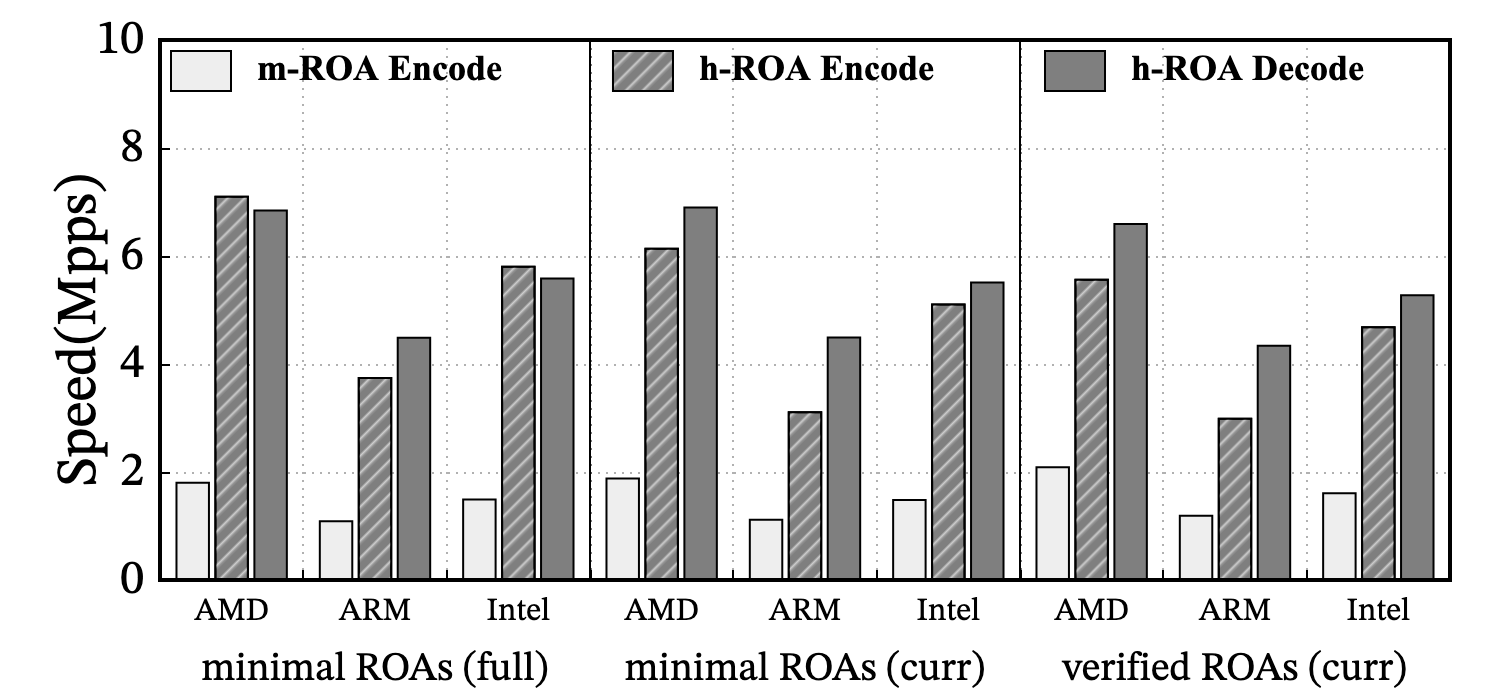}
    \caption{Encoding / Decoding speed in IPv6.}
    \label{fig:speed:v6}
\end{figure}

We evaluate encoding / decoding speeds of \texttt{m-ROA} and \texttt{h-ROA} across three data sets on different platforms, each equipped with a type of CPU. As shown in Fig.~\ref{fig:speed:v4} and~\ref{fig:speed:v6}, though \texttt{h-ROA} brings in additional decoding cost,
its encoding speed is faster than that of \texttt{m-ROA} on all platforms. In other words, \texttt{h-ROA} outperforms \texttt{m-ROA} $1.99 \sim 3.28$ times in terms of the encoding speed.

\subsection{Performance of the reference system implementation}
\label{sect:exp:system}

In the reference implementation (as introduced in Section~\ref{sect:system}), the RP implements the encoding functions of \texttt{m-ROA} and \texttt{h-ROA} respectively, while the router implements the decoding function of \texttt{h-ROA}. \texttt{m-ROA} does not need any changes on the router.
% We evaluate encoding / decoding speeds of both schemes on two cloud servers. As shown in Fig.~\ref{exp:encode:cloud-1} and Fig.\ref{exp:encode:cloud-2}, though \texttt{h-ROA} brings in additional decoding cost on the router, its encoding speed is faster than that of \texttt{m-ROA} on both platforms. In addition, the time spent on decoding hanging ROAs constructed with \texttt{curr} on both platforms are $143$ms and $250$ms respectively, which are only $7.7\%$ and $8.0\%$ of the total time the router spends on synchronization. In another word, the decoding cost can be ignored when taking the whole process in consideration. So can the encoding cost.
% \begin{figure}[!tbp] %cloud1 阿里云& cloud2科技云
%   \begin{minipage}[b]{0.48\linewidth}
%     \centering
%     \includegraphics[width=\textwidth]{results/cloud-1.png}
%     \caption{Encode speed on \texttt{cloud-1}.}\label{exp:encode:cloud-1}
%   \end{minipage}%
%   \hfill
%   \begin{minipage}[b]{0.48\linewidth}
%     \centering
%     \includegraphics[width=\textwidth]{results/cloud-2.png}
%     \caption{Encode speed on \texttt{cloud-2}.}\label{exp:encode:cloud-2}
%   \end{minipage}
% \end{figure}
We measure the synchronization time of PDUs, which is the time elapsed between the time when the router sends an
RTR \texttt{reset} query and that when it receives the \texttt{End of
Data} PDU. Synchronization time integrates the costs on transmission, encoding / decoding and inserting as a whole. Two kinds of system deployment are evaluated.
% This statement is verified with the evaluation on the total time spent in synchronization with an RTR \texttt{reset} query, which integrates the costs on transmission and encoding / decoding as a whole, and is measured as the time elapsed from the router sending a \texttt{reset} query to the time when it receives the \emph{End of Data} PDU. Two kinds of system deployment are evaluated. 

\begin{figure}[!tbp]
    \centering
    \includegraphics[width=\linewidth]{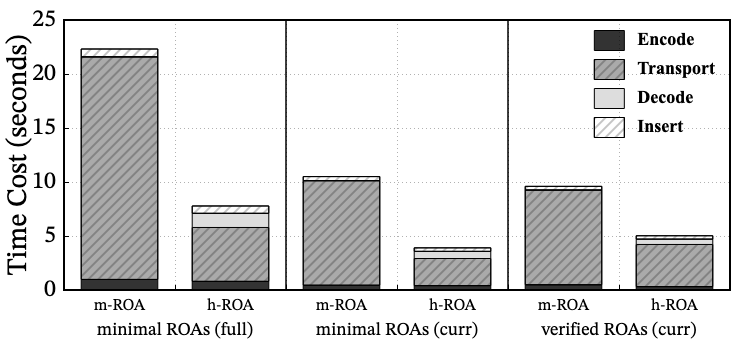}
    \caption{RPstir2-GoBGP (Intra-domain).}
    \label{fig:rpstir2-gobgp:intra}
\end{figure}

\begin{figure}[!tbp]
    \centering
    \includegraphics[width=\linewidth]{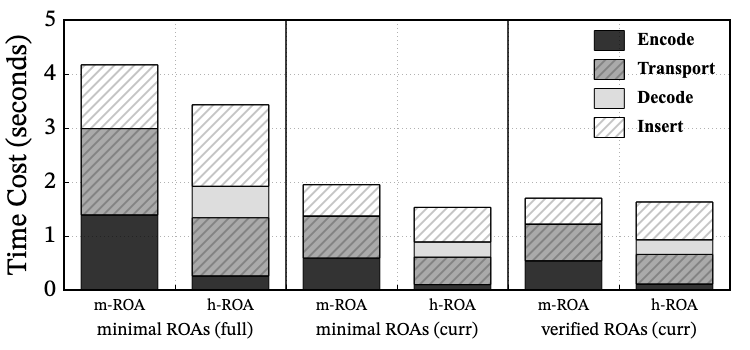}
    \caption{Routinator-FRRouting (Intra-domain).}
    \label{fig:routinator-frr:intra}
\end{figure}

In the first scenario, which is referred as ``intra-domain", the RP and the router are deployed with different servers on the same local area network, minimizing the influence of transmission since the bandwidth between the two ends are maximized. As shown in Fig.~\ref{fig:rpstir2-gobgp:intra} \~{} Fig.~\ref{fig:routinator-frr:intra}, compared with \texttt{m-ROA}, \texttt{h-ROA} reduces the synchronization cost at least by 21.5$\%$ and 4.0$\%$ with the minimal ROAs and verified ROAs in \texttt{curr}, respectively, which turns to be 17.5$\%$ with the minimal ROAs in \texttt{full}. 

\begin{figure}[!tbp]
    \centering
    \includegraphics[width=\linewidth]{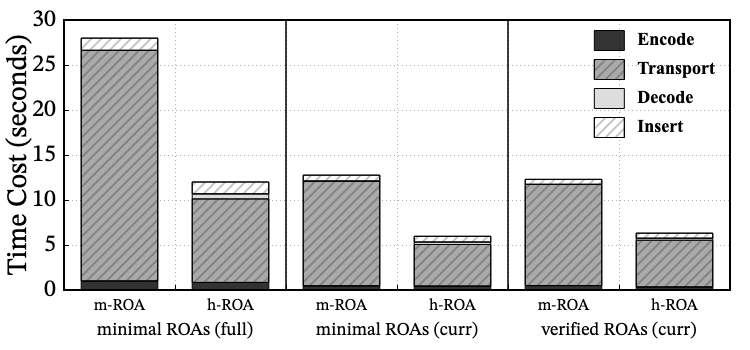}
    \caption{RPstir2-FRRouting (Intra-domain).}
    \label{fig:rpstir2-frr:intra}
\end{figure}

\begin{figure}[!tbp]
    \centering
    \includegraphics[width=\linewidth]{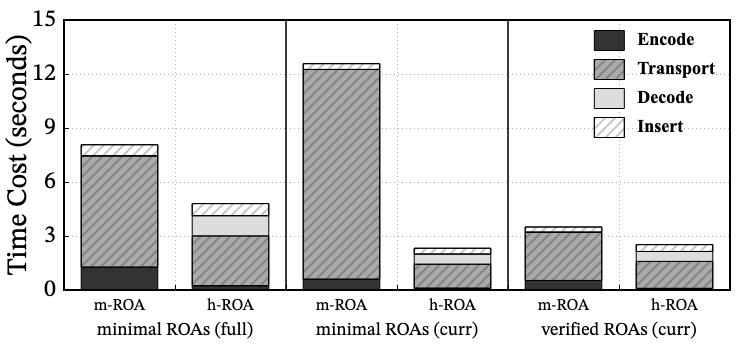}
    \caption{Routinator-GoBGP (Intra-domain).}
    \label{fig:routinator-gobgp:intra}
\end{figure}

\begin{figure}[!tbp]
    \centering
    \includegraphics[width=\linewidth]{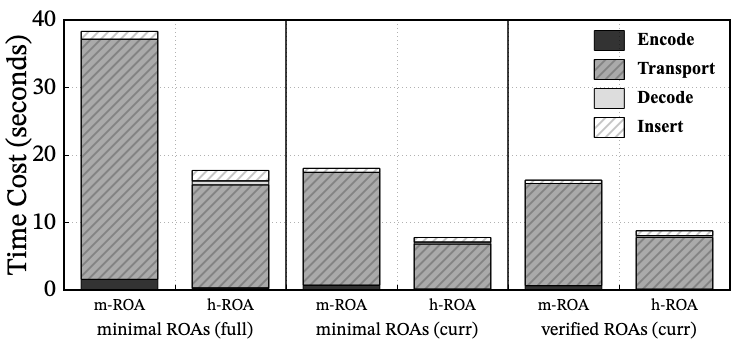}
    \caption{Routinator-FRRouting (Cross-domain).}
    \label{fig:routinator-frr:cross}
\end{figure}

In the second scenario,  which is referred as ``cross-domain", the RP and the router are deployed on different servers at different clouds, where the bandwidth between the two servers is around 10~Mbps. 
%To simulate a more extreme case, we restrict the bandwidth to around 10~Mbps. 
As shown in Fig.~\ref{fig:rpstir2-gobgp:intra} \~{} Fig.~\ref{fig:routinator-frr:cross}, the cross-domain deployment always takes a longer time than that within the same server. Besides, we can see clearly that \texttt{h-ROA} has a higher efficiency in synchronization than \texttt{m-ROA}, and though \texttt{h-ROA} brings in additional decoding cost on the router, its encoding speed is faster than that of \texttt{m-ROA} in all cases with different data sets, on different platforms, with different kinds of system deployment. In another word, the decoding cost can be ignored when taking the whole process in consideration. So can the encoding cost.
%And the superiority of \texttt{h-ROA} turns to be larger when the bandwidth is limited.
\begin{figure}[!tbp]
    \centering
    \includegraphics[width=\linewidth]{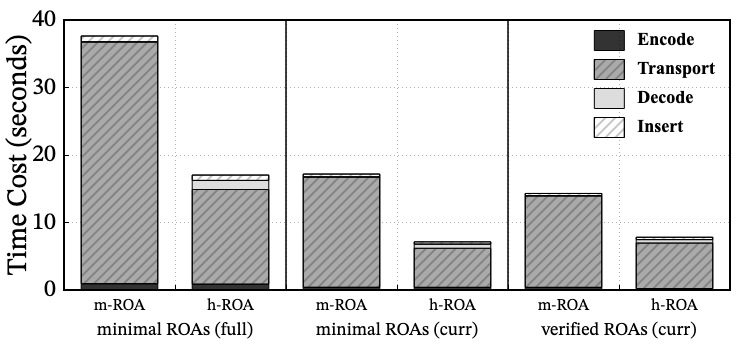}
    \caption{RPstir2-GoBGP (Cross-domain).}
    \label{fig:rpstir2-gobgp:cross}
\end{figure}

\begin{figure}[!tbp]
    \centering
    \includegraphics[width=\linewidth]{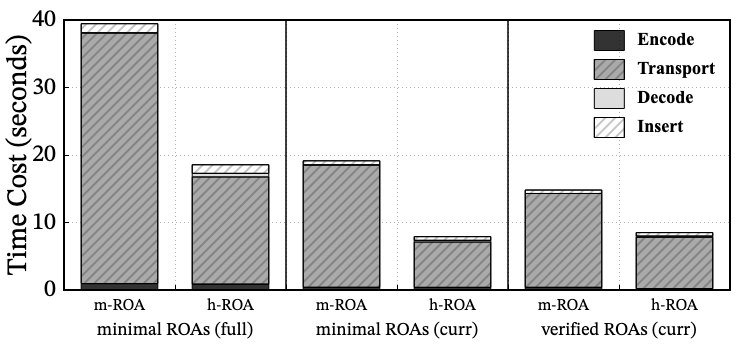}
    \caption{RPstir2-FRRouting (Cross-domain).}
    \label{fig:rpstir2-frr:cross}
\end{figure}

\begin{figure}[!tbp]
    \centering
    \includegraphics[width=\linewidth]{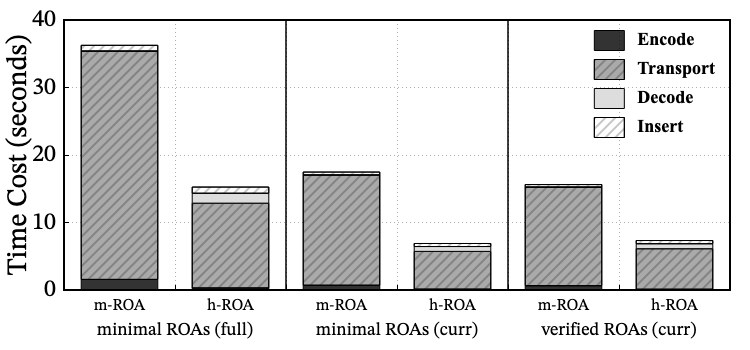}
    \caption{Routinator-GoBGP (Cross-domain).}
    \label{fig:routinator-gobgp:cross}
\end{figure}

% \begin{figure}[!tbp] %域内& 域间
%   \begin{minipage}[b]{0.48\linewidth}
%     \centering
%     \includegraphics[width=\textwidth]{results/sameserver.png}
%     \caption{Deployment on one server.}\label{exp:system:integrated}
%   \end{minipage}%
%   \hfill
%   \begin{minipage}[b]{0.48\linewidth}
%     \centering
%     \includegraphics[width=\textwidth]{results/crossdomain.png}
%     \caption{Cross-domain deployment.}\label{exp:system:cross}
%   \end{minipage}
% \end{figure}
\section{Related Work }%0.5p
\label{sect:work}

RPKI is the infrastructure to secure the inter-domain routing with BGP. Besides the ROV with ROAs, path verification is another hot topic, on which many schemes have been proposed~\cite{rfc8374-bgpsec, one-hop,path-end, aspa-profile, aspa-verification}. With the route origin validated with ROAs, the path verification guarantees the integrity and compliance of the propagation path of a BGP announcement . However, most of them are confronted with great resistances in practical deployment. For example, BGPsec~\cite{rfc8374-bgpsec} poses a heavy burden on routers, and is vulnerable to downgrade attacks~\cite{li-bgpsec}. ASPA~\cite{aspa-profile, aspa-verification} uses sensitive business information to prevent route leaks, which may lead to privacy leakage. To date, ROV with ROAs is the only RPKI-based approach that has been used in practice.

Several studies~\cite{rpki-study,rpki-coming}~\cite{clark2020filter,reuter2018towards,rpki-web} have measured the status of RPKI deployment, including the issuance of ROAs and the adoption of ROV. It has been pointed out that partial deployment of RPKI and ROAs can bring in limited benefits~\cite{forged-origin-attack}. Fortunately, the rate of ROA issuance has increased rapidly in recent years~\cite{rpki-coming}. Regarding a full deployment in the near future, existing mechanisms will suffer from serious challenges in performance and scalability~\cite{rpki-full-size}. \texttt{h-ROA} is designed to address these challenges.
%The contributions we make in this paper are to address these challenges to some extent.

ROAs are useful in stopping prefix hijacks, but their mis-configurations can lead to many issues~\cite{rpki-study,rpki-coming,forged-origin-attack, classification}, damaging the stability and security of the inter-domain routing system. DISCO~\cite{disco} proposes an automated certification system to address the downward dependency~\cite{forged-origin-attack}. It is also suggested to limit the number of prefixes contained in one ROA for higher availability~\cite{multi-prefix-roa}. However, balancing availability and scalability is challenging, where \texttt{h-ROA} can lead to possible solutions with its use of flexible and controllable compression.

\section{Conclusion}%0.25p
\label{sect:end}

In this paper, we present the design of the {\em bitmap-based ROA} (\texttt{BM-ROA} in short), a secure and flexible encoding scheme for ROAs, especially for those scattered authorization cases. Furthermore, we propose the {\em hybrid ROA} (\texttt{h-ROA}), which is a scheme that combines {\em maxLength-based ROA} (\texttt{ML-ROA}) and \texttt{BM-ROA}, dynamically selecting the encoding scheme between them according to the distribution pattern of the authorized IP prefixes of an AS, in order to achieve the maximum compression effect and the most flexible compression management. Then, we evaluate the performance of \texttt{h-ROA} via extensive experiments and a real system implementation. \texttt{h-ROA} is as secure as the state-of-the-art approach, but is more flexible and scalable and outperforms it significantly in compression effects according to our evaluations. In addition, \texttt{h-ROA} can sharply reduce the cost of a router in synchronizing all validated ROA payloads. 

Here are some recommendations on the encoding of ROAs for their use in operational RPKI systems. First of all, the minimal ROA principle should be followed as much as possible to ensure the secure use of ROAs, and the resulted scalability issue can be addressed with \texttt{BM-ROA}. Second, in response to special-purpose ROAs, \texttt{h-ROA} can be a better alternative to the only use of \texttt{maxLength} or \texttt{bitmap} in ROAs, because the degree of compression \texttt{h-ROA} can achieve has outperformed that can be achieved by using only \texttt{maxLength} or \texttt{bitmap}, and it supports more flexible and controllable compression. Last but not the least, it is worthwhile trading the encoding complexity with a deeper compression, because the transmission cost dominates the overall ROA synchronization cost on the router.

%We have provided public access to our code and data at \texttt{\url{https://www.gobeta.ac.cn/rpki/hroa}}.

%\input{8-ack}

% reference
\bibliographystyle{unsrt}
%\bibliography{rpki}

% 重置表格和图片的计数器
\section*{Appendix}

\setcounter{table}{0}
\setcounter{section}{0}
\renewcommand{\thetable}{A.\Roman{table}}
\renewcommand{\thesection}{A.\Roman{section}}
\section{The mappings between the index and the RRDP domain name (CNAME) of 50 publication points}

\clearpage
\onecolumn
\begin{longtable}{|c|c|}
\caption{The mappings between the index and the RRDP domain name (CNAME) of publication points.} \\ \hline
Index & RRDP domain name (CNAME) \\ \hline
\endfirsthead

\caption[]{The mappings between the index and the
RRDP domain name (CNAME) of 65 publication points} \\ \hline
Index & RRDP domain name (CNAME) \\ \hline
\endhead
\hline
\endfoot
\hline
\endlastfoot

\toprule
Index & RRDP domain name (CNAME) \\
\midrule
\endhead
\midrule
%\multicolumn{2}{r}{{Continued on next page}} \\
\midrule
\endfoot

\bottomrule
\endlastfoot
1 & rpki.rand.apnic.net \\
2 & rpki.akrn.net \\
3 & rpki.apnic.net \\
4 & rsync.rpki.tianhai.link \\
5 & rpki-rps.arin.net \\
6 & 0.sb \\
7 & rpkica.twnic.tw \\
8 & rpki-rsync.us-east-2.amazonaws.com \\
9 & rpki-rsync.mnihyc.com \\
10 & rpki.sub.apnic.net \\
11 & rpki.co \\
12 & rpki.roa.net \\
 13 & rpki.owl.net \\
 14 & rpki-repository.nic.ad.jp \\
 15 & rpki.admin.freerangecloud.com \\
 16 & rpki.apernet.io \\
 17 & repo-rpki.idnic.net \\
 18 & rsync.rp.ki \\
 19 & rpki.cnnic.cn \\
 20 & rsync.paas.rpki.ripe.net \\
 21 & ca.nat.moe \\
 22 & krill.accuristechnologies.ca \\
 23 & repo.kagl.me \\
 24 & cloudie-repo.rpki.app \\
 25 & dev.tw \\
 26 & rpki.pedjoeang.group \\
 27 & rpki.multacom.com \\
 28 & rpki.tools.westconnect.ca \\
 29 & rpki.arin.net \\
 30 & rpki-01.pdxnet.uk \\
 31 & rpki.cc \\
 32 & rpki.sailx.co \\
 33 & rpki.afrinic.net \\
 34 & repository.lacnic.net \\
 35 & rpki-repo.registro.br \\
 36 & rpki.netiface.net \\
 37 & rsync.krill.cloud \\
 38 & rpki.xindi.eu \\
 39 & pub.krill.ausra.cloud \\
 40 & rpki.pudu.be \\
 41 & chloe.sobornost.net \\
 42 & rpki.services.vm.i.bm-x0.w420.net \\
 43 & rpki.zappiehost.com \\
 44 & rpki.nap.re \\
 45 & r.magellan.ipxo.com \\
 46 & rpki.komorebi.network \\
 47 & ca.rg.net \\
 48 & rpki.folf.systems \\
 49 & rpki.ripe.net \\
 50 & rpki.sn-p.io \\
\end{longtable}
\twocolumn

\end{document}